\documentclass[
 aps,
 amsmath,amssymb,
 reprint,%
]{revtex4-2}

\usepackage{graphicx}
\usepackage{dcolumn}
\usepackage{bm}
\usepackage{tikz}
\usepackage{hyperref}
\hypersetup{%
    colorlinks=true
}

\begin{document}

\preprint{APS/123-QED}

\title{Magnetocaloric effect in multilayers studied by membrane-based calorimetry}

\author{M. Kulyk}%
\email{mykolak@kth.se}
\altaffiliation[Also at ]{Institute of Physics of the NAS of Ukraine, 46 Nauky Ave., Kyiv 03028, Ukraine}
\author{M. Persson}%
\author{D. Polishchuk}
\altaffiliation[Also at ]{Institute of Magnetism of the NAS of Ukraine and MES of Ukraine, 03142 Kyiv, Ukraine}
\author{V. Korenivski}
\affiliation{Nanostructure Physics, Royal Institute of Technology, 10691 Stockholm, Sweden}
\date{\today}

\begin{abstract}
We study magnetic multilayers, incorporating dilute ferromagnetic spacers between strongly-ferromagnetic layers exhibiting a proximity-enhanced magnetocaloric effect. Using magnetometry and direct measurements of the adiabatic temperature change based on a nanomembrane-calorimetry, we find that the magnetocaloric effect in the studied multilayer is indeed enhanced compared to that in the bulk spacer material. We develop a phenomenological numerical model of the studied trilayer and find that a long-range exchange interaction through the weakly-ferromagnetic spacer is required to adequately describe the magnetic and magnetocaloric properties of the system. 
%
\end{abstract}

\maketitle

\section{\label{sec:level1}Introduction}
Since the discovery of the giant magnetocaloric effect (MCE) in Gd$_5$(Si$_2$Ge$_2$)~\cite{Pecharsky1997Giant}, the literature reveals a steadily growing interest in the area of MCE and magnetic cooling, in particular. The related technology of adiabatic demagnetization refrigeration (ADR) is expected to benefit from the new materials and offer cheaper, robust, more compact and long-lasting solutions than the commonly used gas-based heat pumps, with the ultimate goal of reducing the energy loss and the greenhouse gas emissions~\cite{Yu2003Review,Metz_2005}. Even though some of the currently available MCE materials can be relatively eco-friendly in the long run, they often need expensive, difficult to mine rear-earth elements, which can neutralize their benefits when compared to the gas-based systems.

The majority of the MCE-related literature is on developing new complex alloys and compounds in the bulk form; only a small fraction of the  MCE-related articles are on MCE in films, multilayers, and other low-dimensional structures. In contrast to bulk materials, nanostructures exhibit finite-size effects (due to surfaces and interfaces), which can significantly alter their MCE with respect to that intrinsic to the incorporated materials~\cite{Miller2014Magnetocaloric,Doblas2017Nanostructuring,Mukherjee2009Magnetocaloric,Vorobiov2019Optimization,Barman2020Interface-induced, doi:10.1080/23311916.2015.1050324}. This can yield novel properties and offer benefits that are unavailable with bulk MCE materials. For example, for a spin valve-like structure with a weakly ferromagnetic spacer~\cite{Kravets2012Temperature-controlled, Kravets2014Synthetic, Kravets_2016}, the field as small as a hundred Oersted could be enough to induce significant MCE~\cite{Fraerman2015Magnetocaloric} due to the magnetic proximity effect (\emph{direct} interlayer exchange across the interfaces). Recent works\cite{Vdovichev2018High, Polushkin2019Magnetic, Kuznetsov2020Magnetocaloric, Kuznetsov2021Magnetocaloric} have investigated such multilayered structures using magnetometry and concluded that the MCE is indeed enhanced, however, no direct magneto-caloric measurements are available. In spin valve-type structures with \emph{indirect} interlayer exchange (RKKY), arranged such that its constructive versus destructive interference (on parallel to antiparallel switching of the spin-valve's magnetization) yields a change in the effective field acting on the dilute ferromagnetic spacer layer of the order of tens of Tesla~\cite{Polishchuk2018Giant}. Such field is strong enough to de/magnetize the spacer layer and induce a giant MCE. Another mechanism for enhancing MCE was demonstrated~\cite{Barman2020Interface-induced} using CoFe$_2$O$_4$/La$_{0.7}$Sr$_{0.3}$MnO$_3$ heterostructures and relies on the interfacial strain-induced magnetostructural coupling in the system. The possibility of tuning between positive and negative MCE in Co/Cr superlattices has been reported~\cite{Mukherjee2009Magnetocaloric}. The above mentioned results illustrate how an enhanced MCE can be obtained from tailoring the interlayer interactions in a nanostructure, rather than by changing the intrinsic properties of the constituent materials. Often, the task is reduced to enhancing the effective de/magnetization field via interfacial spin-spin or spin-orbit interactions and focusing it on the MCE-active region of the nanostructure, designed to have high magnetic susceptibility.

One of the promising nanostructured materials expected to have enhanced MCE is a trilayer of two strongly ferromagnetic films separated by a thin weakly ferromagnetic spacer, F$_{\text{p}}$/f/F~\cite{Kravets2012Temperature-controlled, Kravets2014Synthetic, Kravets_2016, Fraerman2015Magnetocaloric}. Here, F$_{\text{p}}$ and F denote a pinned and a free strongly ferromagnetic layers, respectively, with their Curie temperature much higher than the operating temperature range of the MCE material. f is a weakly ferromagnetic spacer of, typically, a transition metal alloy, with the Curie temperature ($T_{\text{C}}^\text{s}$) much lower than that of the F-layers, designed to lie within the desired operating temperature range by tailoring the magnetic dilution of the alloy. The pinning of the F$_{\text{p}}$ layer can be achieved by exchange bias, high coercive field, or any other effect that is capable to strongly pin the magnetization of the layer. Exchange enhanced MCE in this design is due to  the interplay of the magnetic proximity effects from the two F/f interfaces within the nearly paramagnetic spacer when the F-layers are switched between the parallel (P) and antiparallel (AP) magnetization orientations. Such magnetic multilayer systems were investigated previously in view of the influence of the proximity effect on, e.g., the induced magnetic properties of a Pt spacer~\cite{Lim2013Temperature-dependent} and the transport properties of a dilute magnetic semiconductor~\cite{Song2011Proximity}.

Previous work~\cite{Vdovichev2018High, Polushkin2019Magnetic, Kuznetsov2020Magnetocaloric, Kuznetsov2021Magnetocaloric} on experimentally determining the magnitude of the magnetocaloric effect in the F$_{\text{p}}$/f/F system was conducted using an indirect method relying on macroscopic magnetization measurements, with the assumption that the analysis of the $M-H$ curves based on the Maxwell relation applies to the orientational transition of the strongly ferromagnetic outer layers (P-to-AP), namely, that the P-AP switching is affected by the state of the weakly paramagnetic spacer (f). The model took the magnetization throughout all layers to be along one axis (z-axis, collinear with P/AP magnetization states) as well as only nearest neighbor interatomic exchange in the \emph{paramagnetic} spacer. Others~\cite{Camley1987Surface, Camley1988Phase, Camley1989Properties} have modeled the system and obtained a twisted spin state in the spacer in the AP configuration, similar to a domain wall, highly sensitive to temperature due to the low Curie point of f. Additionally, it was shown~\cite{Magnus2016Long-range} that a more accurate description of the switching behavior in a metallic F/f/F trilayer requires a long-range interaction term in the total energy of the system, especially within the dilute ferromagnetic spacer.

The applicability of the thermodynamic Maxwell relation to modeling the magnetization reversal loops of a system of type F$_{\text{p}}$/f/F must be carefully examined as $M-H$ loops obtained using regular magnetometry (such as VSM) represent an orientational change of the field-projection of the total magnetization dominated by the F-layers, whereas the MCE is related to the changes in the magnetization of the spacer, which are most often indistinguishable by magnetometry. Compounding the difficulties with the Maxwell-relation approach, the presence of typically significant anisotropy and coercivity in the magnetization reversal behavior of the strongly ferromagnetic layers must be carefully analyzed and understood as their contribution to the overall $H-M$ response of the system can be predominant. These considerations make the indirect (via magnetometry) methods of evaluating the MCE in the discussed F/f/F system uncertain and prompted us to implement a nanomembrane-based measurement technique for directly measuring the adiabatic temperature change of the multilayer material undergoing P-AP magnetization reversal in an applied magnetic field.

\section{\label{sec:sampleAndMethods}Samples and Methods}
\subsection{\label{ssec:multilayers}Magnetic Multilayers}
The multilayer samples under investigation were fabricated using magnetron sputtering from multiple targets in a UHV chamber  (ATC Orion system by AJA Inc.). The base pressure in the chamber was lower than $1\mathrm{E}{-8}$~Torr. The Ar pressure during the deposition was $3\mathrm{E}{-3}$~Torr. The deposition rates of Py (Fe$_{20}$Ni$_{80}$), Fe, FeMn (Fe$_{50}$Mn$_{50}$), and Cr were calculated from a nanoprofiler-based calibration of the thickness of the films deposited for each target separately. The Fe$_\text{x}$Cr$_\text{100-x}$ (x=30) alloy was deposited by co-sputtering of the Fe and Cr targets. The atomic ratio x in the Fe$_\text{x}$Cr$_\text{100-x}$ alloy was controlled by the ratio of the deposition rates of Fe and Cr. The obtained thickness of the deposited stacks was in good agreement with the calculated one. To select a direction for exchange bias of the pinned layer, a DC magnetic field of 400~Oe was applied to the substrate during the deposition process. During the film growth, to achieve a more uniform deposition, the substrate (with DC magnets) was rotated with a frequency of 32~rpm. 

The spacer was chosen to be FeCr because of a straightforward correlation of the proportion of Fe in Cr with the Curie temperature of the alloy. The value of the MCE in bulk Fe$_{30}$Cr$_{70}$ is relatively low~\cite{RaviKumar2018Thickness}, nevertheless, it was found to be sufficient in strength and optimal methodologically to act as a calibration for the direct MCE-measurement technique used.

\begin{figure}[!t]
\centering
\includegraphics[width=0.3\textwidth]{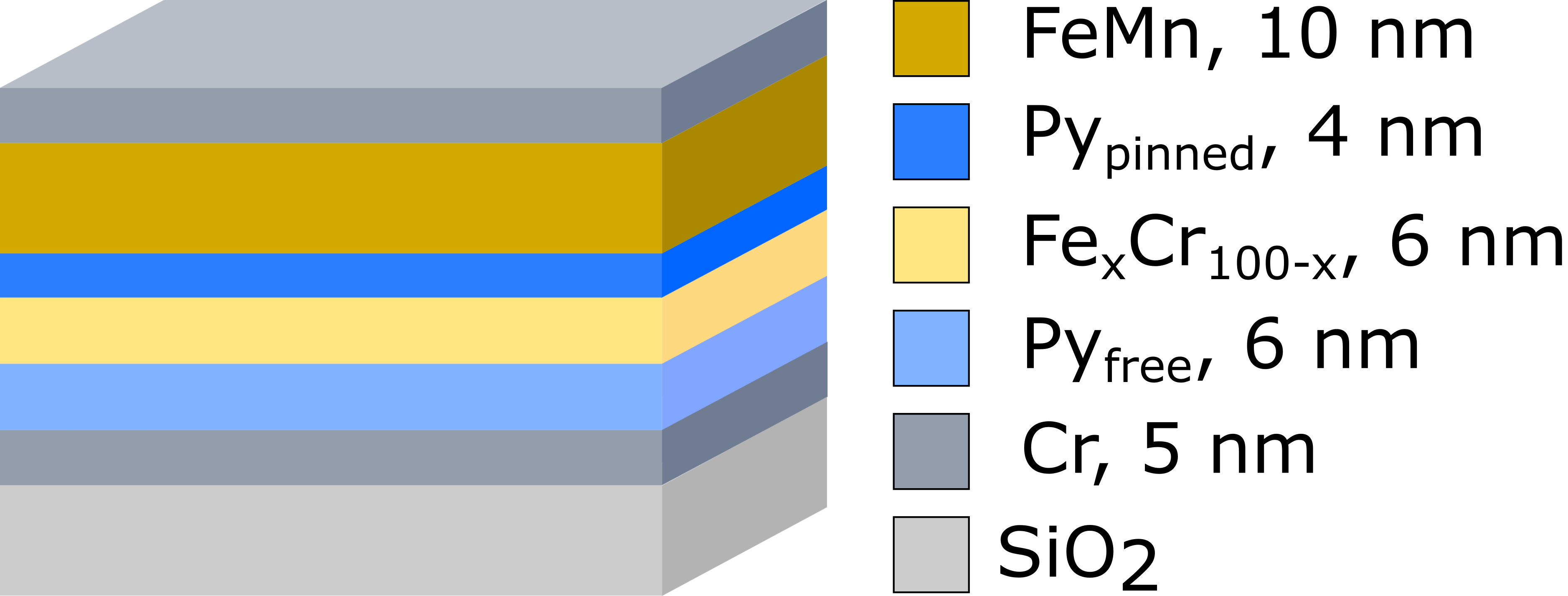}
\caption{\label{fig:layers} Schematic of Py/Fe$_\text{x}$Cr$_\text{100-x}$/Py/FeMn stack.}
\end{figure}

The deposited F$_{\text{p}}$/f/F multilayers have a well-known structure of a spin valve. In a conventional spin valve, the spacer between the pinned and free layers is nonmagnetic, whereas in the system under investigation the spacer is a dilute ferromagnetic alloy with the bulk Curie temperature designed to be close to 160~K. The schematic of the stack is shown in figure~\ref{fig:layers}. To provide better adhesion and improve the growth conditions a 5~nm Cr seed layer was used. Also, a 5~nm Cr layer was used to protect the top of the multilayered film. For a higher signal-to-noise ratio in both vibrating sample magnetometry (VSM) and direct MCE measurements, the Py/Fe$_{30}$Cr$_{70}$/Py/FeMn/Cr structure was sequentially deposited 3 times and the Cr layers served as magnetic isolation between the top FeMn layer and the bottom free Permalloy layer of the next stack.

For this study, we deposited 6 samples, 3 of which onto thermally oxidized Si substrates (``s" series) and 3 onto home-fabricated sensors based on 100~nm SiN thin membranes for direct MCE measurements (``m" series). The naming and description of the samples are shown in table~\ref{tab:tab_1}.

\begin{table*}[!t]
\centering
\caption{\label{tab:tab_1} Multilayer composition and sample notations.}
\begin{tabular}{|c|c|c|}
\hline
Si/SiO substare & SiN membrane & Deposited structure\\
\hline
sA & mA & Cr/Py/Fe$_{30}$Cr$_{70}$/Py/FeMn/Cr\\
\hline
sB & mB & Cr/Py/Cr/Py/FeMn/Cr\\
\hline
sC & mC & Cr/Fe$_{30}$Cr$_{70}$/Cr\\
\hline
\end{tabular}
\end{table*}

The magnetic properties of the multilayers deposited onto oxidized Si substrates should be essentially same as those deposited on membrane-sensors, which are capped with a 15~nm thick insulating layer of SiO$_2$ (sensor details below). To verify this two identical magnetic stacks were deposited onto a bare oxidized-Si substrate and onto a Ti-O layer (sensor material) prepared on the same substrate, where the Ti-O was subsequently capped with a 15~nm of SiO$_2$. VSM $M-H$ measurements for the two samples showed only minor differences in the behavior of the magnetization and, therefore, justified a comparative analysis magnetometry-vs-MCE presented below.

Samples sA and mA were the focus of the study, expected to show an enhanced MCE compared to that in sample mC with the reference value of MCE of the bulk spacer material. The thickness of the Fe$_{30}$Cr$_{70}$ alloy film in samples sC and mC was 50~nm and the $T_{\text{C}}$ close to that of the bulk material~\cite{RaviKumar2015Magnetic}. We, therefore, in what follows refer to samples sC and mC as \emph{bulk} samples.

Samples sB and mB have the same structure as sA and mA but a nonmagnetic spacer, which at 6~nm thickness should not mediate any significant magnetic exchange (direct or indirect) between the free and pinned strongly ferromagnetic layers. This sample is used as an additional control sample in the direct measurements of the adiabatic temperature change of the multilayers, as it should not show any signal related to the MCE in the spacer (Cr spacer is non-magnetic), while it should reveal any spurious contributions that may be present in a membrane-based MCE measurement.

\subsection{\label{ssec:sensors}Sensor Fabrication}
\begin{figure*}[t]
\centerline{
\begin{tikzpicture}[font=\sffamily]
\begin{scope}
\node (a) at (0,0) {\includegraphics[scale=0.13]{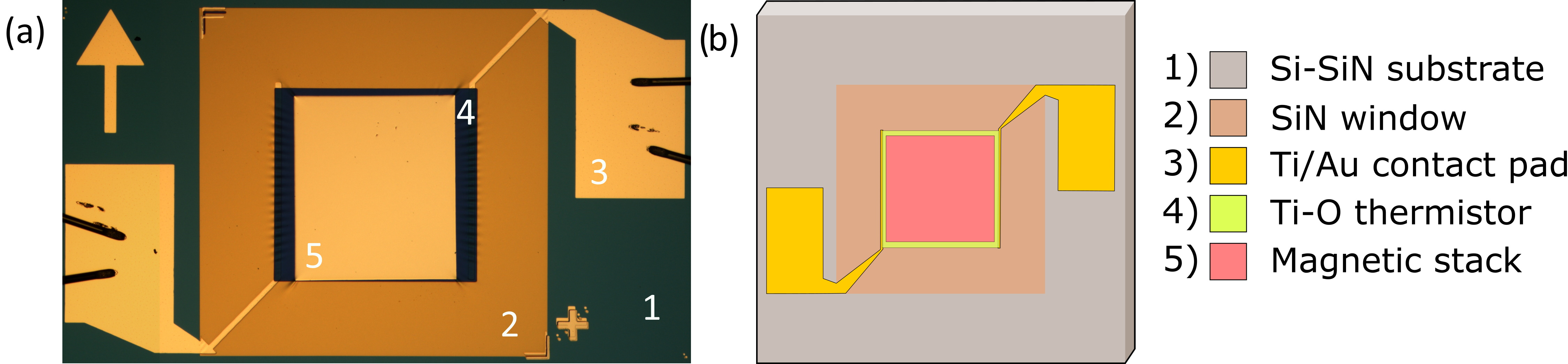}};
\end{scope}

\end{tikzpicture}
}
\caption{\label{fig:sensor}Microphotograph (a) and schematic (b) of sample on Ti-O sensor, on SiN membrane, with following bottom-to-top layer sequence: (1) Si-SiN substrate (7.5~mm $\times$ 7.5~mm $\times$ 381~\textmu m); (2) SiN membrane window (1.5~mm $\times$ 1.5~mm $\times$ 100~nm); (3) contact pads of thermistor with thickness 60~nm (50~nm of Ti, 10~nm of Au); (4) Ti-O planar thermistor (880~\textmu m $\times$ 840~\textmu m $\times$ 50~nm) capped with SiO$_2$ insulation layer (15~nm); (5) magnetic stack (700~\textmu m $times$ 800~\textmu m). Dark lines on contact pads in (a) are bonded wires.}
\end{figure*}

To detect the adiabatic temperature change of the magnetic multilayer a micro-sensor was fabricated consisting of a 50~nm thick Ti-O planar thermistor with high temperature coefficient of resistance (TCR=1-3\%) deposited on a 100~nm SiN membrane (Silson Ltd, UK). The thermistors were deposited using reactive magnetron sputtering from a Ti target in a mixed atmosphere of Ar and O$_2$. The thickness control of the Ti and SiO$_2$ was done in the same way as for the magnetic multilayer. To electrically isolate the thermistor from the conductive magnetic film, a 15~nm layer of SiO$_2$ was deposited on top of the Ti-O layer. The key properties of nm-thin SiN membranes are their low heat conductivity and heat capacity, which make the MCE measurement conditions close to adiabatic~\cite{Tagliati2009Membrane-based, Herwaarden2005Overview}. The micrograph and schematic of the membrane with the thermistor and the magnetic multilayer are shown in figure~\ref{fig:sensor}.

\subsection{\label{ssec:vsm}VSM Measurements}
The VSM measurements were carried out using a Lake Shore 7300 vibrating sample magnetometer equipped with a nitrogen gas-flow cryostat. The isothermal magnetization reversal loops were measured with the field applied in the plane of the films, with the exchange bias direction oriented along the positive direction of the field. The measured magnetization was normalized by the magnetic volume of the samples.

\subsection{\label{ssec:directMeas}Membrane-based Calorimetry}
\begin{figure}[!t]
\centering
\includegraphics[width=0.25\textwidth]{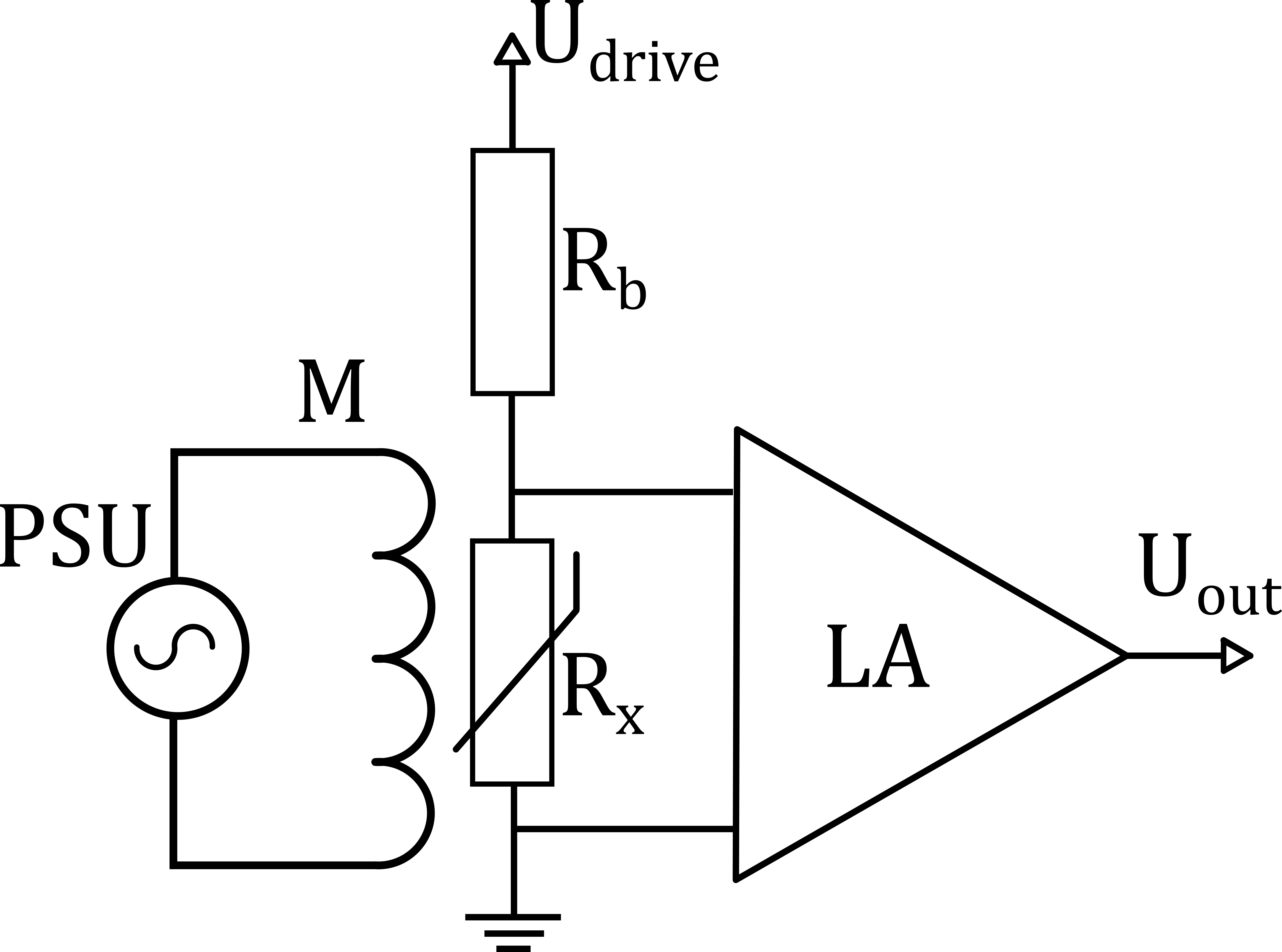}
\caption{\label{fig:setup} Schematic of electrical circuit for measuring resistance of membrane thermistor (R$_\text{x}$, see figure~\ref{fig:sensor}), calibrated to yield temperature change of membrane with sample.}
\end{figure}

The schematic of the measurement setup is shown in figure~\ref{fig:setup}. R$_\text{x}$ denotes the Ti-O planar thermistor, deposited directly onto the SiN membrane, acts as the substrate for deposition of the magnetic multilayer to be studied. DC voltage U$_\text{drive}$ is applied to voltage divider R$_\text{b}$-R$_\text{x}$, where R$_\text{b}$ is a current limiting resistor. The voltage drop across R$_\text{x}$ is fed into a lock-in amplifier (LA). The membrane with the sensor and magnetic stack is placed in a vacuum inside a cryostat (Oxford OptistatDN, modified to operate without gas in the sample space to prevent parasitic sample-gas-chamber heat exchange). The cryostat is placed inside an electromagnet (M), which is powered by a power supply unit (PSU) capable of supplying an AC+DC current (Kepco BOP). 

When a low frequency (11.7 Hz) AC magnetic field is applied to a magnetic sample, sufficient to induce MCE (strong enough to produce P-AP switching in spin-valve samples), the temperature of the sample placed in direct contact with the sensor/membrane changes synchronously with the external field. Due to the low heat conductivity of the membrane, a momentary change in its temperature equilibrates over a relatively long period of time, $\tau>$0.1 s in our case (time constant of our membrane sensor). Thus, for the applied field frequency that exceeds 1/$\tau$ the condition for the temperature change of the sample to be adiabatic is fulfilled (it changes/oscillates faster than it can dissipate to the bigger substrate via the membrane). The magnetic sample is in direct contact with the thermistor, whose resistance R$_\text{x}$ oscillates producing a variation in the voltage drop across it, which is then detected using a lock-in amplifier synced to the frequency of the applied AC magnetic field. Generally, depending on the value of the DC field offset and the properties of the magnetic system, the MCE signal can be detected at the first or the second harmonic. In what follows, we limit the discussion to the most straightforward case of a low-amplitude AC field applied at various fixed biasing fields, such that the signal is found predominantly at the 1$^\textrm{st}$ harmonic (details below).

The size of the planar thermistor must fulfill two main criteria: the volume of the active MCE material must be relatively large to provide a significant change in temperature of the membrane/sensor/trilayer stack; the sensor/trilayer must be relatively small to avoid the unwanted heat exchange with the bulk Si substrate. We have modelled these aspects numerically, that for a given membrane geometry, AC-field frequency, and material parameters, and arrived at the optimal size of the sensor -- the one used in this article.   

As the MCE-related signal is detected at the first harmonic of the AC magnetic field, the inductive pick-up is detected together with the MCE signal. It is unwanted and is removed using the phasor approach to the signal processing. Two measurements are carried out consecutively - with positive and negative U$_\text{drive}$ applied to the R$_\text{b}$-R$_\text{x}$ voltage divider. Denoting the vector of the parasitic inductive pick-up as $\textbf{P}$, the vectors of the MCE signal as $\textbf{S}_1$ and $\textbf{S}_2$ (for positive and negative U$_\text{drive}$), and the vectors measured by lock-in signals as $\textbf{L}_1$ and $\textbf{L}_2$, we have
\begin{equation}
\left\{\begin{matrix}
{\textbf{S}}_{1}=-{\textbf{S}}_{2}={\textbf{S}}_\text{MCE}\\ 
{\textbf{P}}+{\textbf{S}}_{1}={\textbf{L}}_{1}\\
{\textbf{P}}+{\textbf{S}}_{2}={\textbf{L}}_{2}

\end{matrix}\right.
\end{equation}
The solution of this system gives us
\begin{equation}
\begin{split}
S_{\text{MCE}} &= | \textbf{S}_1| = \frac{1}{2}\sqrt{R_1^2 + R_2^2 - 2R_1R_2\cos(\theta_1 - \theta_2)}\ ,\\
\theta_{\text{MCE}} &= \text{atan2}\left(R_1\cos\theta_1-R_2\cos\theta_2, R_1\sin\theta_1 - R_2\sin\theta_2\right)\ ,
\end{split}
\end{equation}
where $R_1$, $R_2$, $S_{\text{MCE}}$ and $\theta_1$, $\theta_2$, $\theta_{\text{MCE}}$ are $R$ and $\theta$ components of the measured and the MCE signal phase vectors, $\textbf{L}_1$ and $\textbf{L}_2$, $\textbf{S}_{\text{MCE}}$, respectively.

The AC component of the resistance of the sensor ($R_\text{AC}$) is equal to
\begin{widetext}
\begin{equation}
\begin{split}
R_{\text{AC}} &= \frac{\big(R_{\text{DC}}R_{\text{I}}+R_{\text{b}}(R_{\text{DC}} + R_{\text{I}})\big)^2S_{\text{MCE}}}{-R_{\text{DC}}R_{\text{I}}^2U_{\text{drive}} + (R_{\text{b}} + R_{\text{I}})\big(R_{\text{DC}}R_{\text{I}} + R_{\text{b}}(R_{\text{DC}} + R_{\text{I}})\big)S_{\text{MCE}}}.
\end{split}
\end{equation}
\end{widetext}
Here $R_{\text{DC}}$ and $R_{\text{I}}$ are the DC resistance of the sensor at a given temperature and the input impedance of the lock-in amplifier (or preamplifier, if used).

For small changes in the thermistor resistance due to the AC-field induced MCE (our case), the temperature coefficient of resistance ($k_{\text{TCR}}$) of the sensor at a given temperature can be considered constant, hence the adiabatic temperature change becomes
\begin{equation}
    \Delta T_\text{ad}^\text{measured} = \frac{-R_{\text{AC}}100\%}{k_{\text{TCR}}R_{\text{DC}}}\ .
    \end{equation}
The heat capacity of the active MCE material is often comparable to or less than that of the sensor-membrane material, a scaling coefficient for $\Delta T_\text{ad}$ needs to be introduced to reflect the ratio of the MCE-active thermal mass to the total thermal mass of the device. Straightforward general considerations show that this thermal-mass correction coefficient ($k_{\text{A}}$) and the resulting $\Delta T_\text{ad}^{\text{MCE}}$ of the magnetic sample only should be equal to
\begin{equation}
k_{\text{A}} = \frac{C_{\text{tot}}}{C_{\text{MCE}}},\quad \Delta T_\text{ad}^{\text{MCE}}=\Delta T_\text{ad}^{\text{measured}}k_{\text{A}} ,
\end{equation}
where $C_{\text{tot}}$ is the total heat capacity of the device including the membrane, the sensor with the SiO$_2$ insulation, the MCE material, and the other auxiliary materials not contributing to the exchange-enhanced MCE in the weakly ferromagnetic spacer (Cr underlayer, biasing antiferromagnet, etc); $C_{\text{MCE}}$ is the heat capacity of the active MCE material only (Fe$_{30}$Cr$_{70}$ in our case); $\Delta T_\text{ad}^{\text{measured}}$ is the raw value of the adiabatic temperature change measured directly; $\Delta T_\text{ad}^{\text{MCE}}$ is the normalized value of the adiabatic temperature change in the MCE material. All values of $\Delta T_\text{ad}$ discussed below are  scaled by $k_{\text{A}}$ to show the characteristics of the studied magnetic material rather than a combined property skewed by the unimportant details of the measurement device (its `dead' thermal mass).

\section{\label{sec:results}Results}
\subsection{\label{ssec:vsm}VSM Measurements}
\begin{figure*}[!t]
\centerline{
\begin{tikzpicture}[font=\sffamily]
\begin{scope}
\node (a) at (0,0) {\includegraphics[scale=0.28]{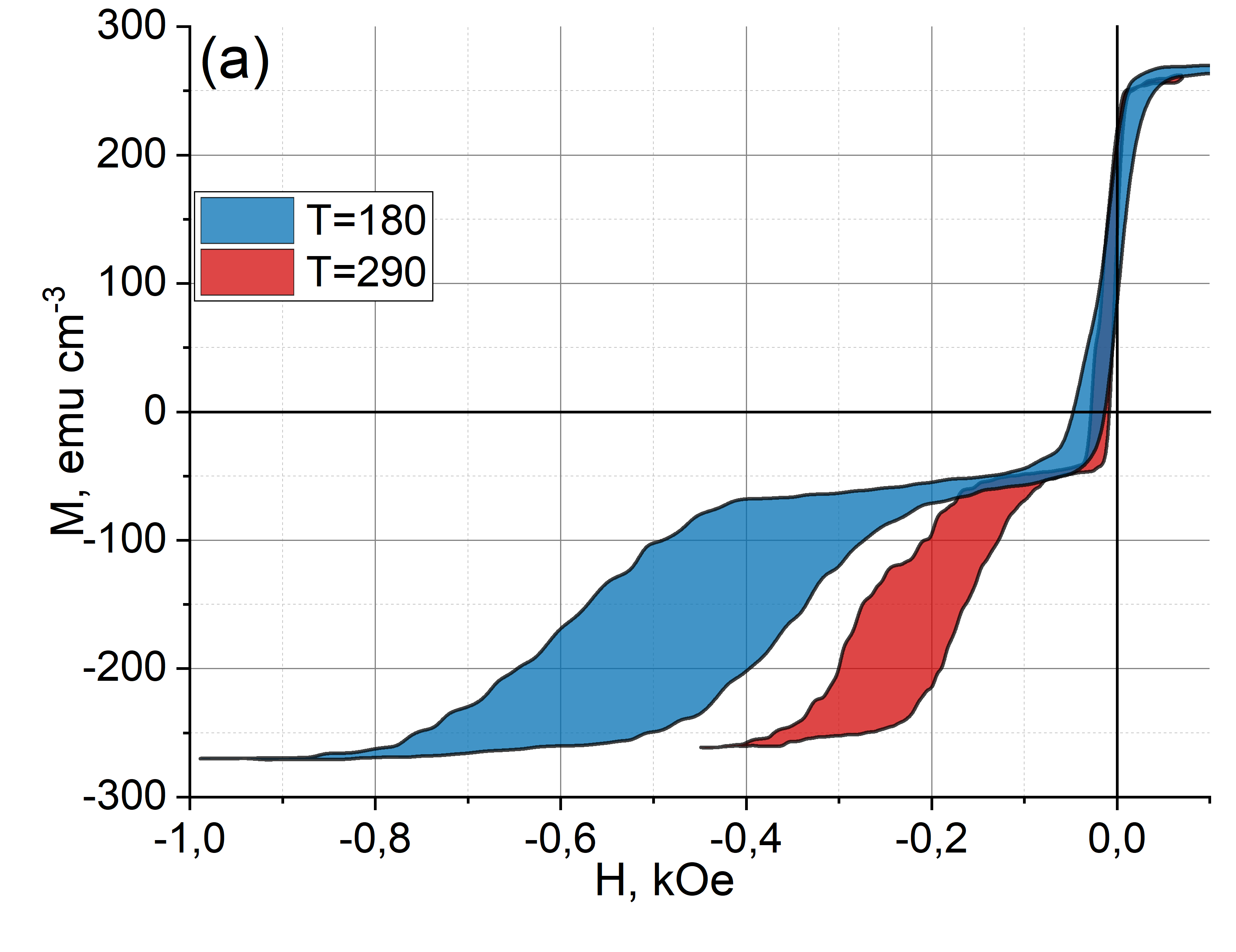}};
\node (b) at (8,0) {\includegraphics[scale=0.28]{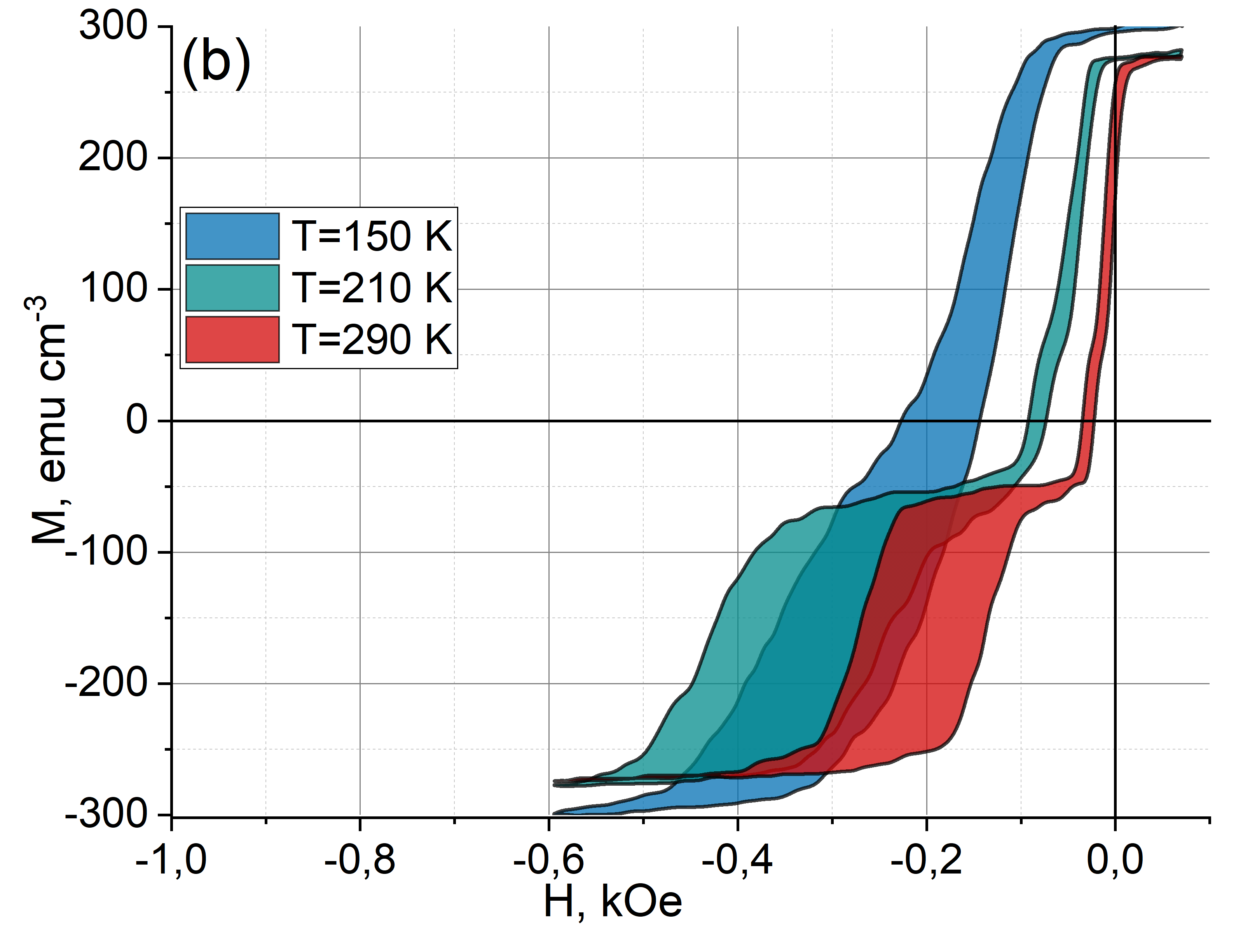}};
\node (c) at (0,-5.8) {\includegraphics[scale=0.28]{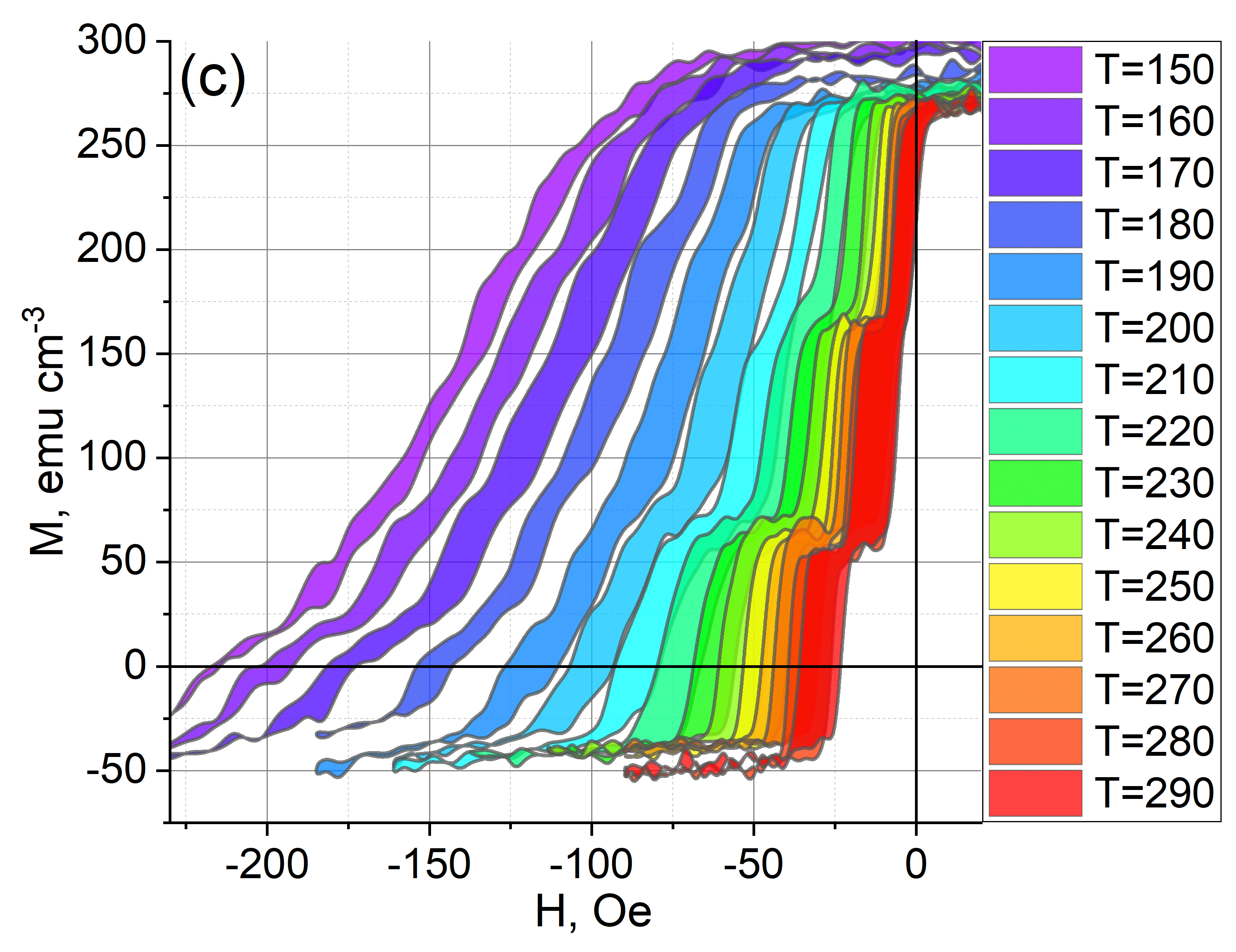}};
\node (d) at (8,-5.8) {\includegraphics[scale=0.28]{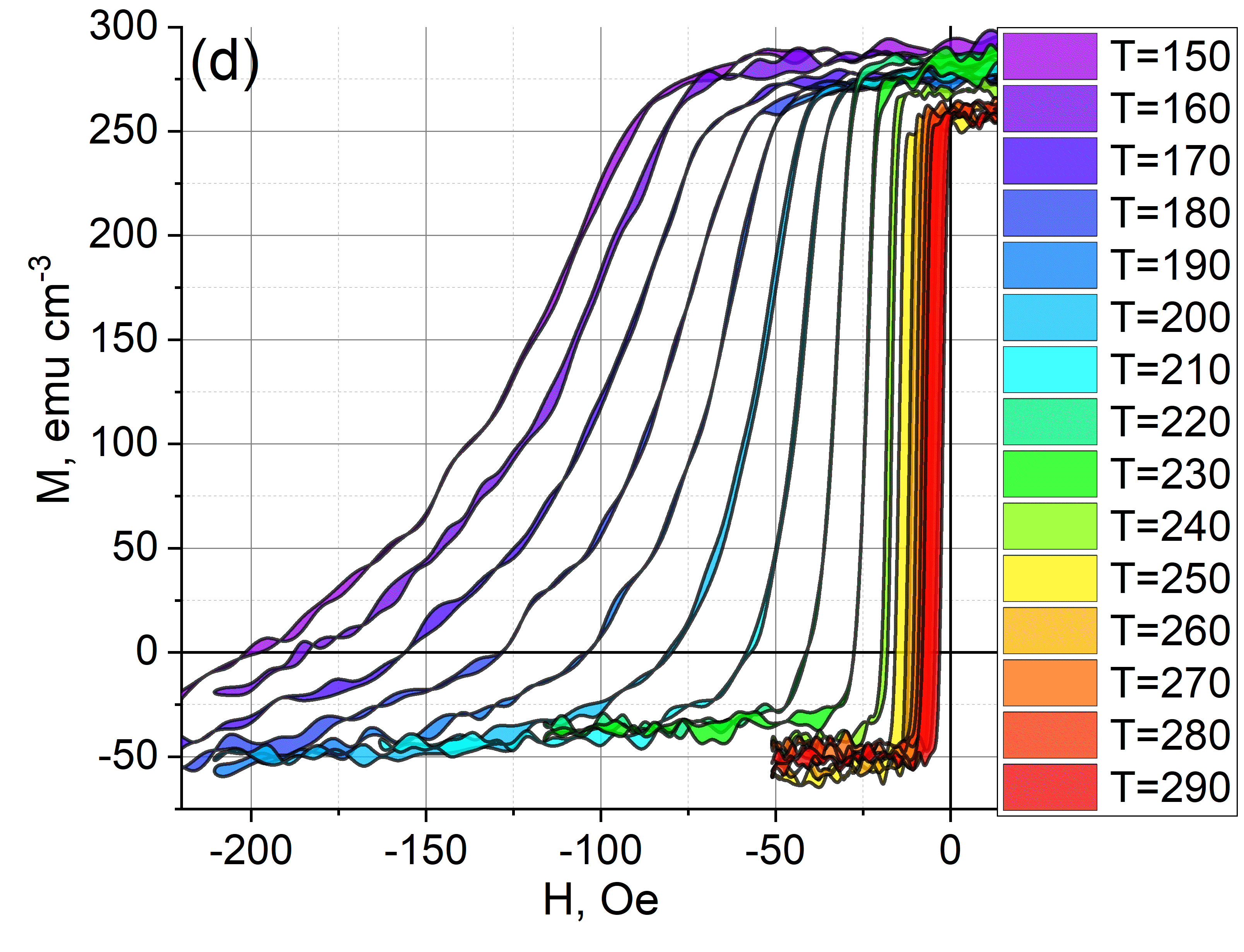}};
\end{scope}

\end{tikzpicture}
}
\caption{\label{fig:4loops}Magnetization reversal loops of samples sB (a), sA (b,c), and single repetition of trilayer of type sA (d).}
\end{figure*}

The magnetization loops for samples sA and sB measured at different temperatures are shown in figure~\ref{fig:4loops}.

Sample sB is a control sample, a spin valve with a nonmagnetic spacer, for which the magnetization reversal loops at $T$=180 and $T$=290 K are shown in figure~\ref{fig:4loops}(a). The behavior is classical behavior for a spin valve~\cite{GoodmanIEEE} -- the hysteresis is composed of two characteristic minor loops. One of the minor loops has low coercivity and is centered at zero field, easily identifiable as switching of the free Py layer. The second minor loop has a higher coercivity and is offset in field, which corresponds to the exchange-biased transition of the FeMn-pinned Py layer. As expected, cooling from 290 to 180 K leaves the minor loop of the free Permalloy layer almost unchanged, while the offset of the pinned loop significantly increases due to a stronger exchange-bias from the antiferromagnetic FeMn at low temperature. The amplitudes of the free and pinned layers' minor loops have a ratio of 6:4, which corresponds to the thickness ratio of the respective layers.

The magnetic properties of the sA sample are similar to those observed in~\cite{Kravets2012Temperature-controlled,Kravets2014Synthetic,Kuznetsov2020Magnetocaloric,Vdovichev2018High}. The $M-H$ loops for three representative temperatures are shown in figure~\ref{fig:4loops}(b). At high temperatures, when the spacer is in its nominally paramagnetic state, the free and pinned layers are almost fully decoupled and their minor loops are detached from each other. Approaching the effective $T_\text{C}^\text{s}$ from above, the exchange coupling through the weakly ferromagnetic spacer increases, which mediates unidirectional anisotropy from the pinned layer and results in a visibly offset loop of the free layer (partially free at 210~K).

\begin{figure}[t]
\centering
\includegraphics[width=0.42\textwidth]{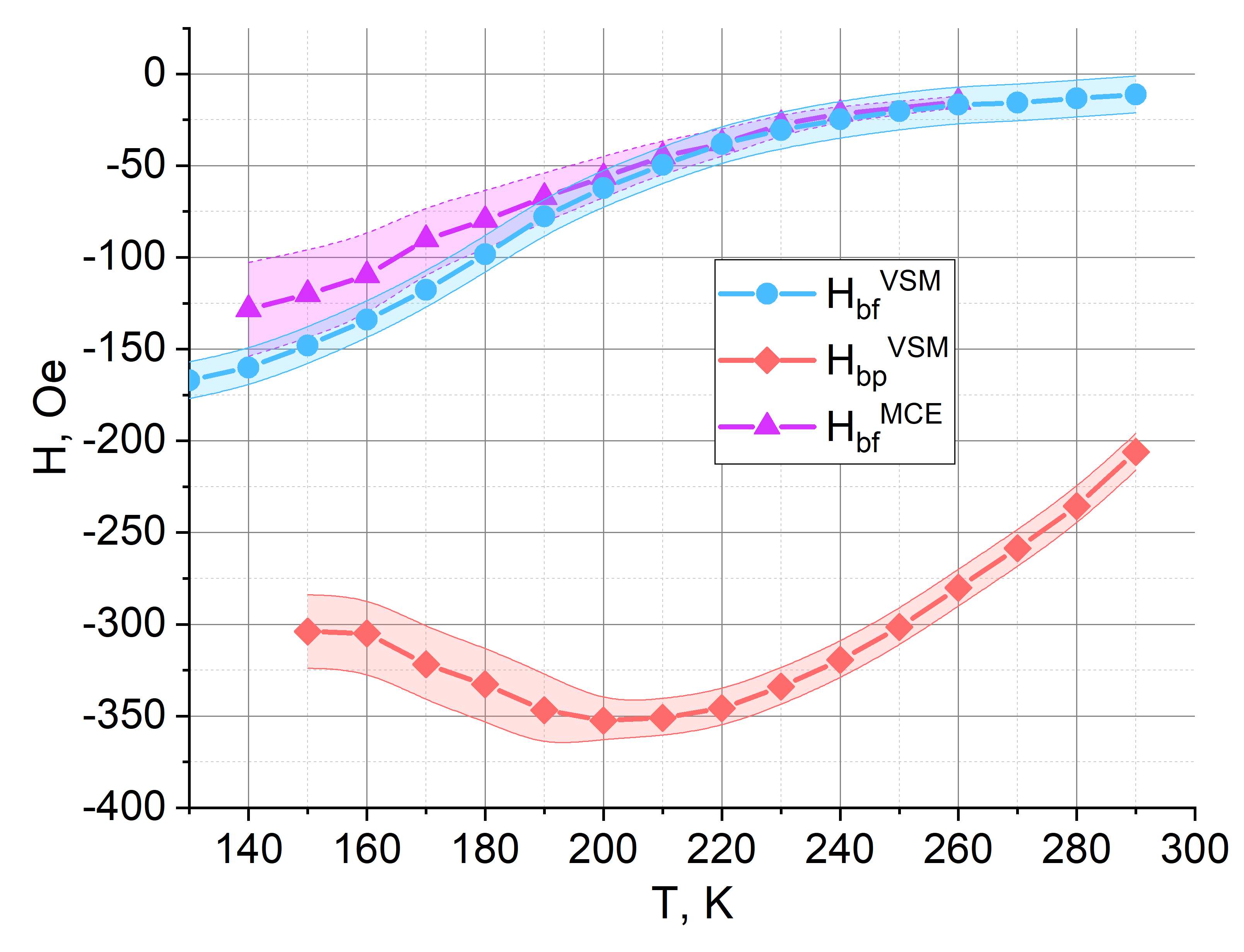}
\caption{\label{fig:Hbias} Temperature dependence of bias field of free layer ($\text{H}_\text{bf}^\text{VSM}$) obtained from $M$-$H$ loops of sample sA (blue circles); bias field of free layer ($\text{H}_\text{bf}^\text{MCE}$) obtained from direct measurements of adiabatic temperature change for sample mA (purple triangles); bias field of pinned layer ($\text{H}_\text{bp}^\text{VSM}$) obtained from $M$-$H$ loops of sample sA (red diamonds).}
\end{figure}

The temperature dependence of the bias field for the free and pinned layer loops for sample sA are shown in figure~\ref{fig:Hbias}. The strength of the biasing field of the free layer increases from essentially zero at high temperature to about 170~Oe at 130~K. The bias-field strength of the pinned layer shows an increase from room temperature (RT) down to about 210~K, after which it weakens on further lowering the temperatures. This non-monotonic behavior is due to an interplay of two effects: first, the well-known increase and eventual saturation of the strength of the antiferromagnetic pinning at low temperature~\cite{Ledue2014Temperature}; the second is related to a higher Zeeman torque, proportional to the effective thickness of the now strongly coupled Py layers at low-$T$, which counteracts the exchange bias and lowers the loop offset~\cite{Ledue2014Temperature,Nogues1999Exchange}. The resulting functional form of $H_\text{bp}^\text{VSM}$ versus $T$ is non-monotonic, as shown in figure~\ref{fig:Hbias} (red curve).

From the low-temperature (150~K) magnetization reversal curve in figure~\ref{fig:4loops}(b) it may appear that the switching of the free layer is a viscous process accompanied by significant coercivity. This conclusion would however not be correct as the partial reversal of only the free layer's magnetization plotted in figure~\ref{fig:4loops}(c) show very low coercivity. The relatively high coercivity of the full $M-H$ loop in the field region of the switching of the free layer is due to the overlap of the narrow switching loop of the free layer and the wide switching loop of the pinned layer.

Comparing the partial magnetization reversal loops of the free layer at room temperature for the samples with the triple and single base spin-valve structure (Figs.~\ref{fig:4loops}(c) and~\ref{fig:4loops}(d), respectively) suggests that, in addition to the magnetic state/dilution of the spacer, the strength of the interlayer coupling depends on the roughness of the interfaces, which increases somewhat for thicker films. It is visible that for the second and third repetitions of the deposited spin-valve structure, the RT exchange bias of the free layer gradually increases (figure~\ref{fig:4loops}(c)).

\subsection{\label{ssec:direct} Direct Measurements of MCE}
\begin{figure*}[t]
\centerline{

\begin{tikzpicture}[font=\sffamily]
\begin{scope}
\node (a) at (0,0) {\includegraphics[scale=0.28]{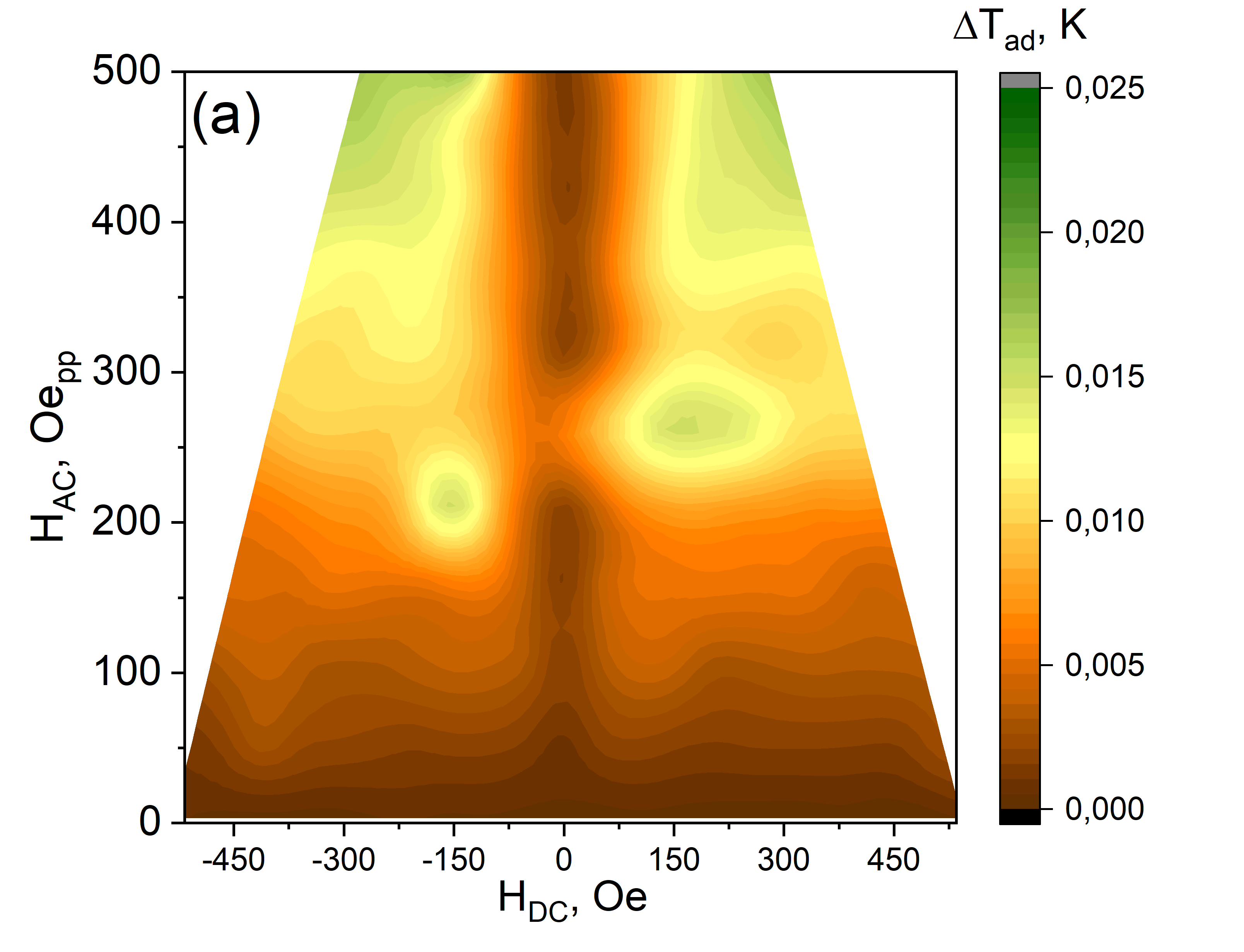}};
\node (b) at (8,0) {\includegraphics[scale=0.28]{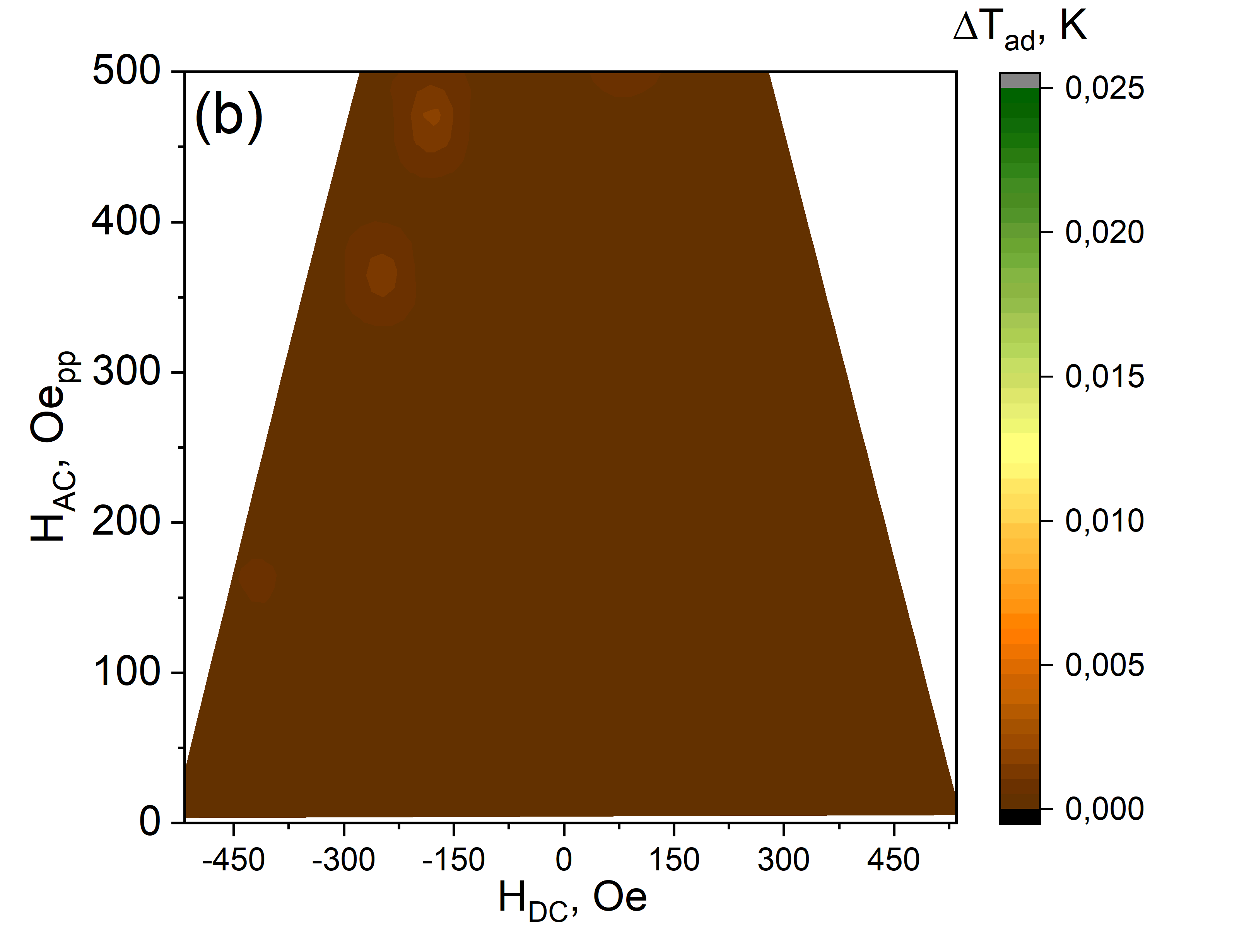}};
\node (c) at (0,-5.8) {\includegraphics[scale=0.28]{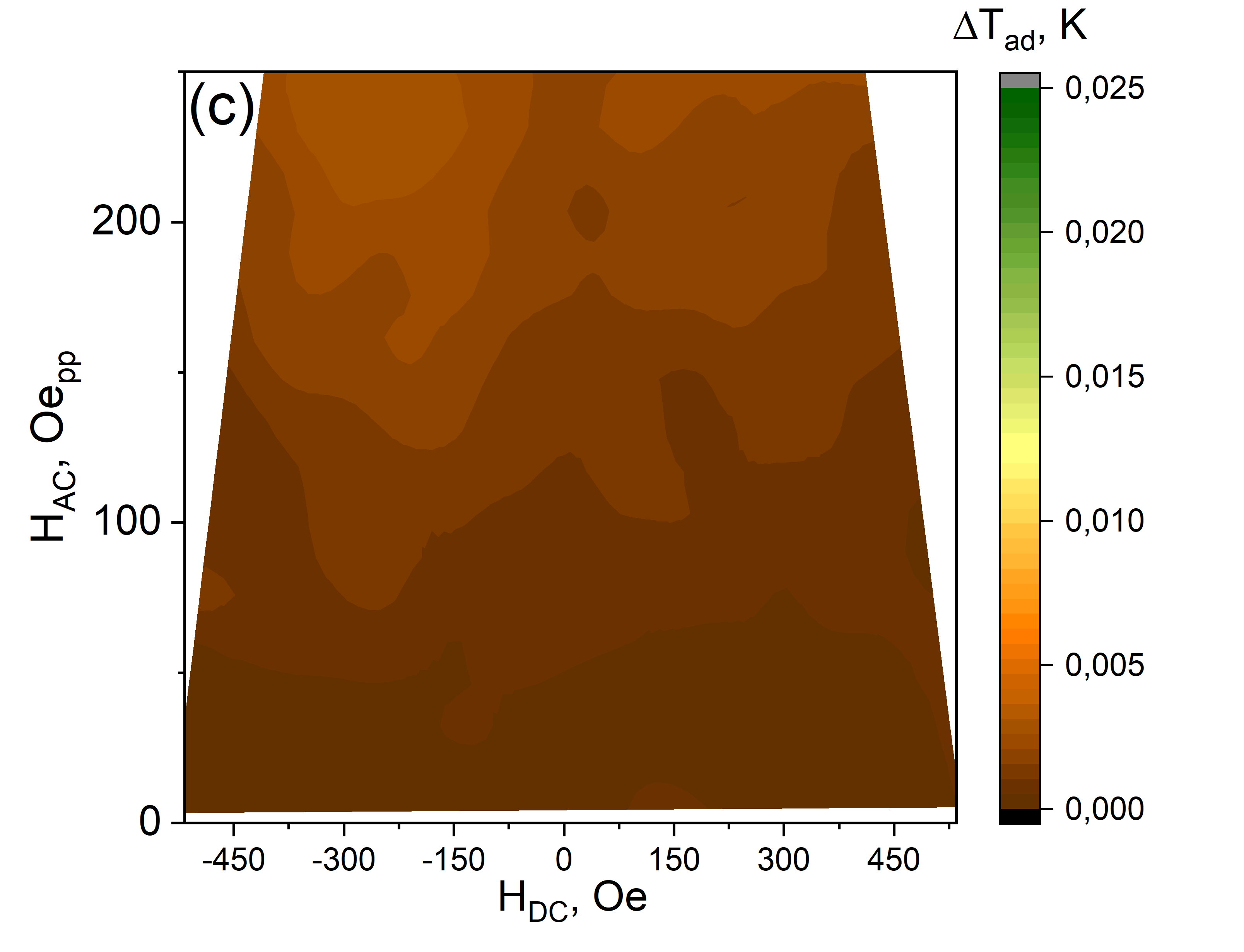}};
\node (d) at (8,-5.8) {\includegraphics[scale=0.28]{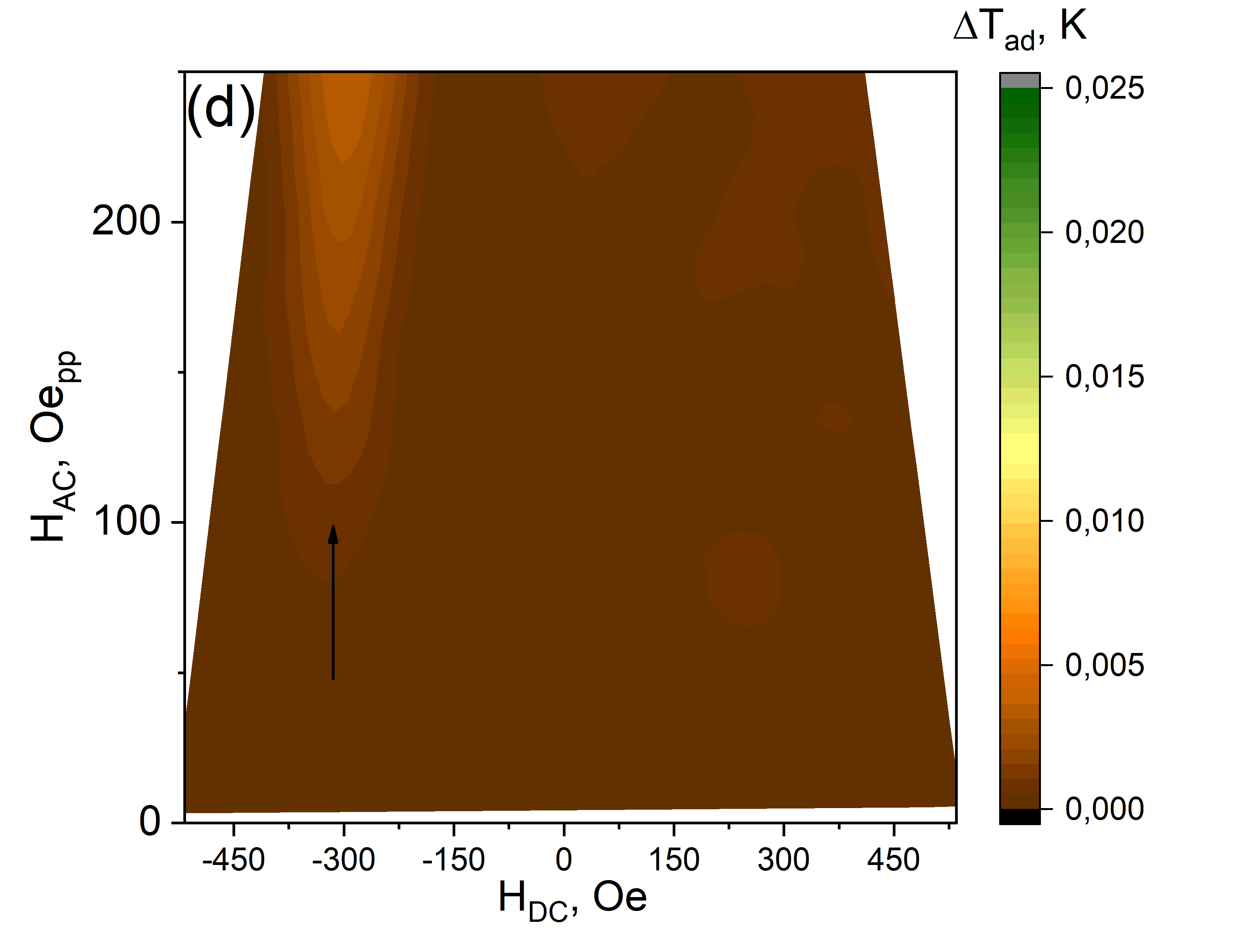}};
\node (c) at (0,-11.6) {\includegraphics[scale=0.28]{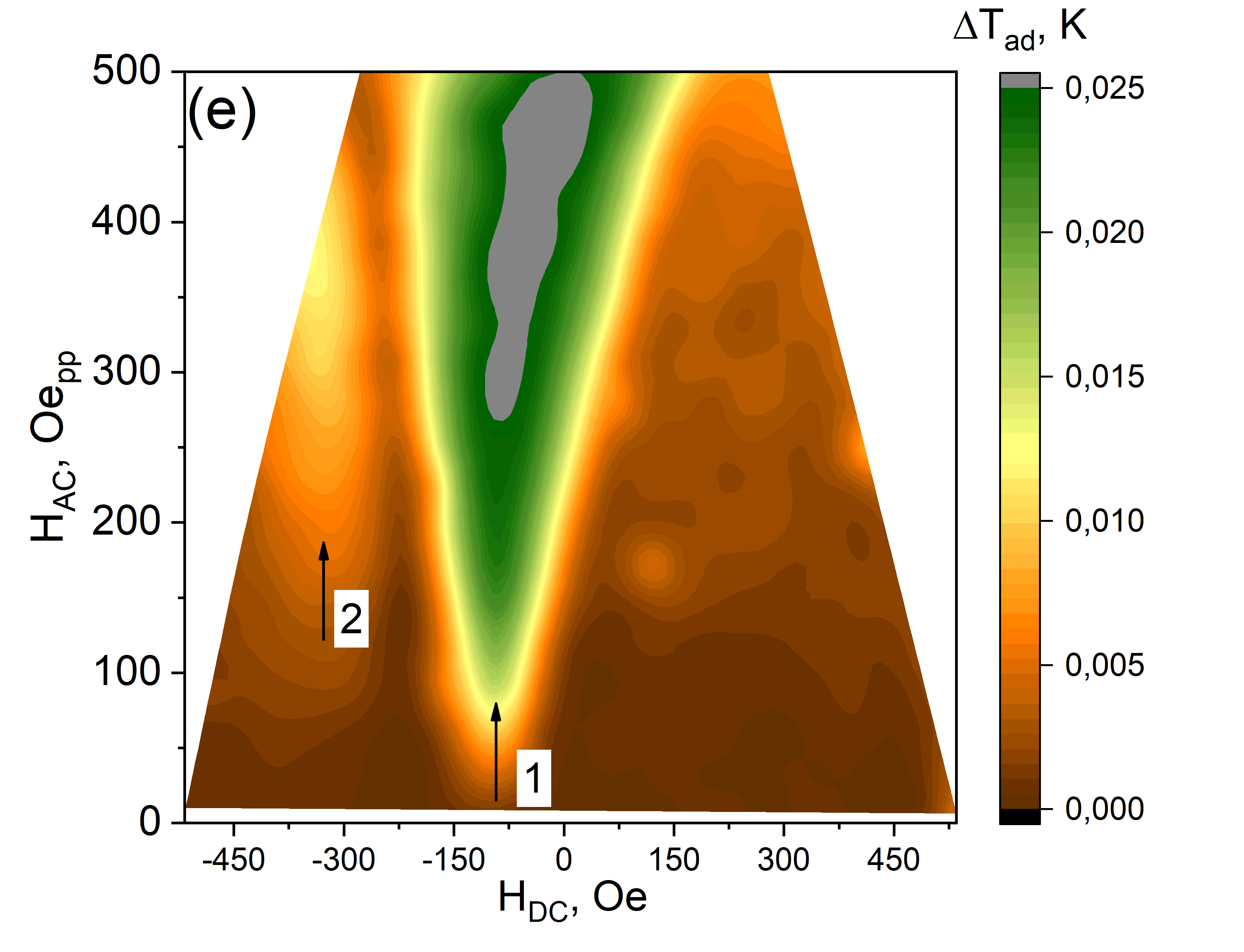}};
\node (d) at (8,-11.6) {\includegraphics[scale=0.28]{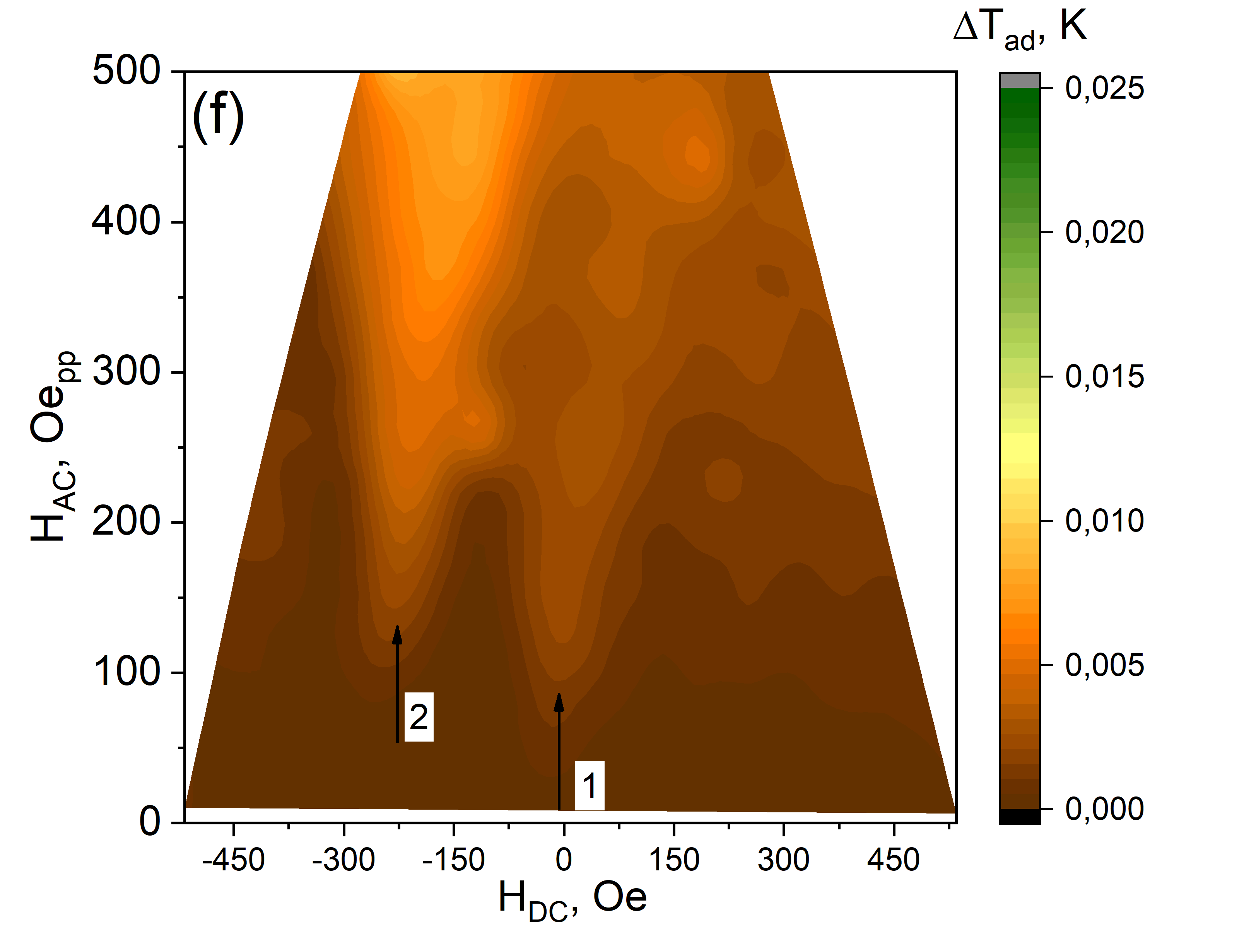}};
\end{scope}

\end{tikzpicture}
}
\caption{\label{fig:MCEmaps}Typical $\Delta T_{ad}$ maps for samples mA (e, f), mB (c, d), and mC (a, b) at following temperatures: (a) – 160 K; (b) – 295 K; (c) – 150 K; (d) – 295 K; (e) – 170 K; (f) – 295 K. Horizontal axis represents DC bias field in Oe; vertical axis --  peak-to-peak value of AC field modulation.}
\end{figure*}

Using the membrane-based MCE device described above we have measured the magnitude of the temperature variation caused by the AC modulation of the field applied to the sample. Typical 2D maps of the MCE signal versus peak-to-peak AC field and DC offset field for samples mA, mB, and mC are shown in figure~\ref{fig:MCEmaps}.

The data for the 50~nm thick film of Fe$_{30}$Cr$_{70}$ alloy (mC sample) are shown in figure~\ref{fig:MCEmaps}(a) and (b) for 150 and 295~K, respectively. At the high temperature, no signal is visible above the background noise. As the temperature is lowered, a clear signal appears and increases, reaching a maximum at $\sim$160~K. The typical appearance of the MCE signal shown in figure~\ref{fig:MCEmaps}(a) remains the same at all temperatures where the signal is detected, changing only in intensity. The gap near zero DC field can be explained by the fact that at $|H_\text{DC}|<\frac{1}{2} H_\text{AC}^\text{pp}$ the value of the field-dependent entropy change $S(H_\text{DC}+\frac{1}{2} H_\text{AC}^\text{pp})- S(H_\text{DC}-\frac{1}{2} H_\text{AC}^\text{pp})$ is lower than for the case of $|H_\text{DC}|>\frac{1}{2} H_\text{AC}^\text{pp}$, as $S(H)$ is an even function and becomes zero at $H_\text{DC}=0$. 

The sample with a spin-valve structure and a nonmagnetic spacer (mB) shows a relatively weak peak at 295~K (figure~\ref{fig:MCEmaps}(d), marked with arrow), localized at the bias field of the pinned layer, determined independently from the respective magnetization reversal loop in figure~\ref{fig:4loops}(a), for the same temperature. At lower temperatures, this peak broadens and shifts to a higher negative DC field (not shown), eventually (toward 170~K) moving out of the experimentally available field range (both AC and DC). This well defined peak clearly visible at 295~K is heating/cooling due to the reversal of the pinned Py layer, which is likely demagnetizing/frustrating a portion of the spins at the F/AF interface, some of which are uncompensated~\cite{PhysRevB.102.140404}. This MCE of the F/AF interface is outside the focus of the present study and will not be discussed further, however, it is an interesting topic in itself and deserves a separate investigation. 

A background signal linearly proportional to the strength of the AC field is detected at lower temperatures. It is DC-field unspecific and, therefore, does not complicate the interpretation of MCE peaks related to the F-layer magnetization switching, which is found at specific DC fields.

For sample mA, two well-distinguishable peaks are observed in the temperature range 150-290~K. At room temperature the first peak (marked with arrow~1 in figure~\ref{fig:MCEmaps}(f)) is located near zero field and has a somewhat lower strength (less bright) than the second peak (arrow 2). Additionally, the signal for peak 1 appears and saturates at a lower AC modulation field than that for peak~2. At low temperatures, both peaks shift to higher negative DC fields and increase in both magnitude and width. The data for $T=170$~K corresponding to the maximum MCE we have observed is shown in figure~\ref{fig:MCEmaps}(e). At temperatures lower than 170~K both signals undergo a rapid change, both in magnitude and shape - they merge and become hardly distinguishable.

We identify peaks 1 and 2 as corresponding to the P-to-AP and AP-to-P switching of the free and pinned layers in the sample (mA -- trilayer with dilute ferromagnetic spacer). Peak~1 corresponds to the transition where the magnetization of the pinned layer stays unchanged while that of the free layer reverses. Peak~2, then, is where the switching occurs in the pinned layer while the free layer is fixed by the relatively high DC field offset. This interpretation is supported by the independently measured $M-H$ switching fields (see magnetometry data above), and naturally explains why the AC field required to induce peak~2 is higher than that for peak~1 -- the width of the partial $M-H$ loop of the pinned layer is much larger than that of the free layer.  

It should be mentioned that peak~2 in figure~\ref{fig:MCEmaps}(f) at high temperatures possibly contains two magnetocaloric contributions -- one from the proximity-enhanced MCE in the paramagnetic spacer, described in the previous paragraph, and one from the MCE that can be related to the F/AF interface and is observed in figure~\ref{fig:MCEmaps}(d). As the amplitude of the latter is much lower and is observed only at high temperatures, we can neglect its contribution to the most prominent MCE regime depicted in figure~\ref{fig:MCEmaps}(e), clearly identifiable as due to the relatively narrow P-to-AP transition on switching of the free layer at about 50-100~Oe in DC offset.

\begin{figure}[t]
\centering
\includegraphics[width=0.5\textwidth]{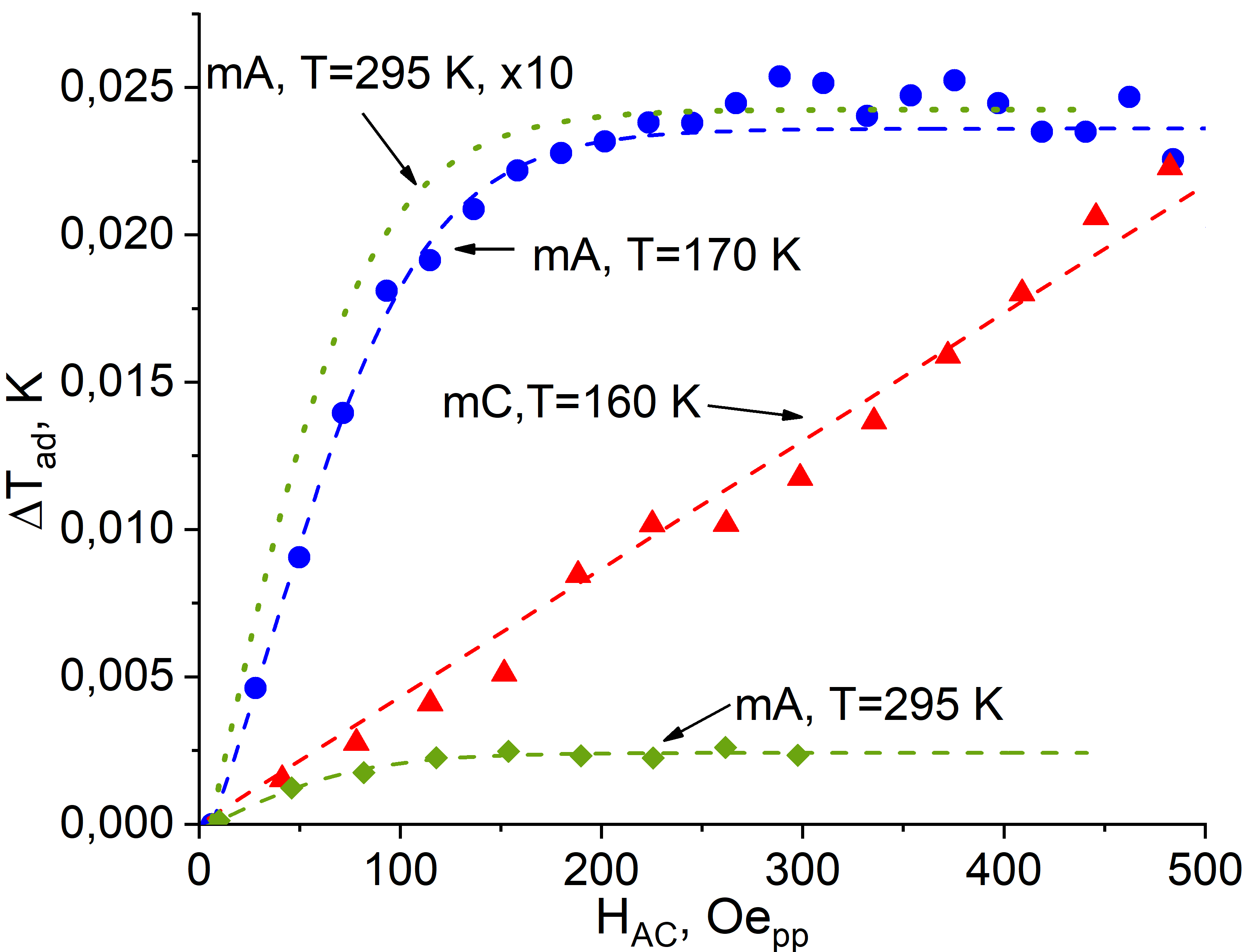}
\caption{\label{fig:slopes} Profiles of $\Delta T_\text{ad}$ vs AC field peak-to-peak amplitude, experimental points with best fits. Blue circles and green diamonds: $\Delta T_\text{ad}$ for sample mA at 170 and 295~K at DC bias of maximum signal; 295~K data scaled up tenfold are shown as green dotted curve to demonstrate same functional form seen in 170~K data. Red triangles: average $\Delta T_\text{ad}$ for sample mC at 160~K at DC bias fields $|H_\text{DC}|>\frac{1}{2} H_\text{AC}^\text{pp}$.}
\end{figure}

The dependence of $\Delta T_\text{ad}$ on $H_\text{AC}$ for samples mA and mC at the temperature of the maximum MCE signal are shown in figure~\ref{fig:slopes}. Since the $\Delta T_\text{ad}$ signal for sample mC is delocalized in DC offset, the profile presented in figure~\ref{fig:slopes} for this sample is the average value of $\Delta T_\text{ad}$ for $|H_\text{DC}|>\frac{1}{2} H_\text{AC}^\text{pp}$. For sample mA, the data presented in figure~\ref{fig:slopes} is the profile of the map in figure~\ref{fig:MCEmaps}(e) corresponding to the center of peak 1. The dependence for sample mC is linear with a positive slope, precisely as expected for a bulk-like alloy. At the same time for sample mA, the behavior of $\Delta T_\text{ad}$ vs AC-field is clearly nonlinear, with a region of rapid growth followed by saturation. We denote the value of $\Delta T_\text{ad}$ at saturation for sample mA as $\Delta T_\text{ad}^\text{s}$.

The temperature dependence of the bias field of the directly measured MCE in sample mA (figure~\ref{fig:Hbias} - purple curve) and the bias field of the partial M-H loop of the free layer in sample sA measured by VSM (figure~\ref{fig:Hbias} - blue curve) are essentially identical. This makes us conclude that the strongest MCE in the F$_{\text{p}}$/f/F structure is manifest on switching of the free layer, producing the P-to-AP transition in the system, with the weakly ferromagnetic spacer more frustrated in terms of its spatial spin orientation (higher magnetic disorder/entropy) between the oppositely magnetized interfaces of the outer F-layers in the AP state.

\begin{figure}[t]
\centering
\includegraphics[width=0.5\textwidth]{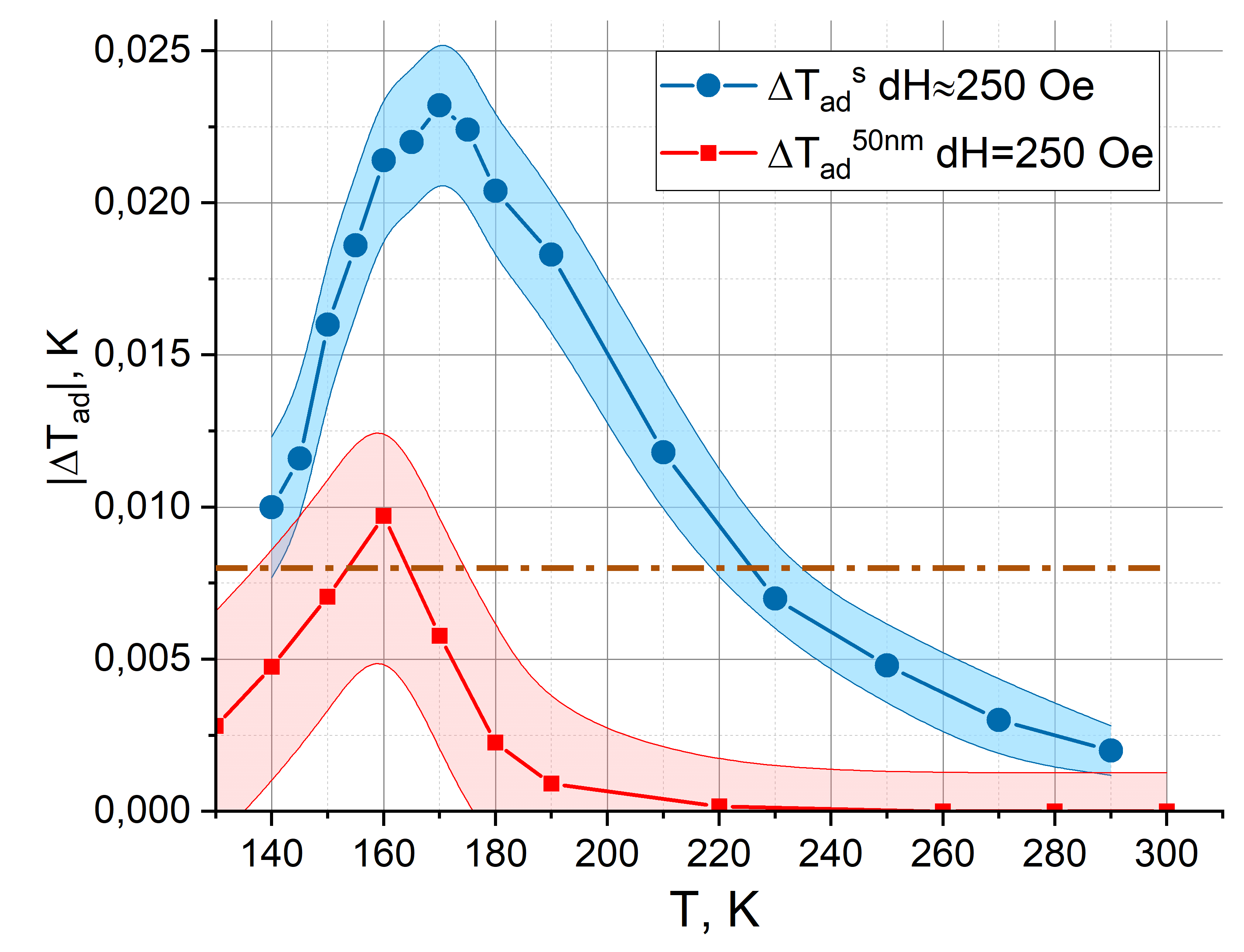}
\caption{\label{fig:Tadcomparison} Adiabatic temperature change obtained from direct MCE measurements: blue circles for spin valve with Fe$_{30}$Cr$_{70}$ spacer, red squares for 50~nm film of Fe$_{30}$Cr$_{70}$, both for $\Delta H=$250~Oe; brown dash-dotted line was obtained using Maxwell's equation for 88~nm of Fe$_{30}$Cr$_{70}$~\cite{RaviKumar2018Thickness}.}
\end{figure}

Figure~\ref{fig:Tadcomparison} shows the data for the adiabatic temperature change measured directly on the membrane-fabricated samples, as well as that reported in the literature for the same composition bulk-like Fe-Cr alloy and extracted from magnetometry using the Maxwell relation~\cite{RaviKumar2018Thickness} at the temperature of maximum MCE near 160~K. For samples mA and mC, MCE maps similar to those shown in figure~\ref{fig:MCEmaps} were recorded for the temperature range of 130-295~K. Using the data in figure~4(a) of Ref.~\cite{RaviKumar2018Thickness} for $\Delta S$ for a 88~nm Fe$_{30}$Cr$_{70}$ film, the value of $\Delta S=0.024$~J/kg K is obtained for the field changing from 0 to 250~Oe. Assuming a constant value of the heat capacity, C=462~J/kg K, the maximum adiabatic temperature change in this case of the bulk-like alloy is $|\Delta T_{ad}|\approx 0.008$~K at $\sim 160$~K (brown dash-dotted line in figure~\ref{fig:Tadcomparison}). The value of the peak adiabatic temperature change we measure directly for the same alloy (our 50~nm thick film of Fe-Cr; sample mC) and the one from the literature obtained indirectly are in good agreement. The Curie temperature for the Fe$_{30}$Cr$_{70}$ sample estimated using the Arrot plots for the VSM M-H data taken at different temperatures for sample sC, is close to 160~K, which is in good agreement with the position of the maximum entropy change in our direct measurement as well as the literature~\cite{RaviKumar2018Thickness,RaviKumar2015Magnetic}.

The $\Delta T_\text{ad}^\text{s}$ vs T dependence for sample mA (figure~\ref{fig:Tadcomparison} blue circles) shows a wide asymmetric peak with a maximum at $\sim$170~K which is $\sim$10~K higher than the Curie temperature of the Fe$_{30}$Cr$_{70}$ alloy. The maximum measured MCE in the spin valve with the dilute ferromagnetic spacer is $\sim$2.5 greater than the intrinsic MCE of the spacer material (taken here for the same external AC field amplitude). Further, the respective half-widths of the MCE signal are $\sim$50~K versus 25~K, which means that, overall, the relative cooling power per unit volume is approximately 5 times higher for the F$_\text{p}$/f/F trilayer~\footnote{We note that using the indirect Maxwell-relation approach with the magnetization data for the trilayer sample yielded MCE, which did not correlate with the direct measurements either in magnitude or temperature dependence, and was discarded from further discussion and analysis.}.

Even though, at room temperature, the 6~nm Fe$_{30}$Cr$_{70}$ spacer is in a nearly paramagnetic state and the coupling between the two outer F$_{\text{p}}$ and F layers should be weak-to-absent~\cite{Kravets2014Synthetic}, both the exchange bias of the F layer (nominally free) and the MCE signal of the trilayer are nonzero. These observations are rather fascinating since the correlation length of the ferromagnetic order parameter is nominally less than 1~nm for Fe$_{30}$Cr$_{70}$ (at RT), suggesting that the direct interatomic exchange should not be expected to provide the observed finite exchange coupling through the 6~nm thick spacer at a temperature that exceed the spacer's $T_\text{C}^\text{s}$ by nearly a factor of two.

An important characteristic of the investigated effect conclusively following from the data of figure~\ref{fig:slopes} is the saturation of the MCE signal with the applied AC field amplitude for sample mA. This means that the magnitude of the MCE in the structure (almost entirely within the weakly magnetic spacer; see Discussion below) predominantly depends on the mutual orientations of the outer strongly ferromagnetic layers and not on the strength of the applied magnetic field, as is the case for the bulk-like Fe-Cr alloy film with a linear MCE-vs-Field dependence.

\section{\label{sec:discussion}Discussion}
The switching of the free layer and the corresponding $\Delta T_\text{ad}$ signal in the studied system (sA and mA samples) are characteristically offset in field even at room temperature, which indicates a non-zero interlayer exchange coupling through the spacer, nominally too thick to mediate direct exchange as the spacer should be in a paramagnetic state, much above its Curie point. The presence of such exchange coupling at temperatures much higher than $T_\text{C}^\text{s}$ in similar F/f/F structures was observed by several groups previously~\cite{Kravets2012Temperature-controlled,Lim2013Temperature-dependent,Kravets2014Synthetic,Magnus2016Long-range,Vdovichev2018High}. It has been suggested~\cite{Magnus2016Long-range} that a long-range magnetic interaction (much longer than the usual inter-atomic exchange length scale) should be added to the energetics of the system in order to properly account for the observed interlayer coupling between the outer ferromagnetic layers at high temperatures and for thick (vs atomic scale) spacers.

A possible origin of such long-range coupling through a metallic spacer is exchange mediated by the polarized conduction electrons~\cite{PhysRevB.54.9353, Gorobets2000} that traverse the stack and can transfer spin over longer distances. The spin flip length ($l_\text{sf}$) in magnetic metals, which is the key characteristic in this process, varies from a few to a few tens of nanometers~\cite{Bass2007Spin-diffusion}. It is then reasonable to expect that in the dilute ferromagnetic metal-alloy spacer electrons polarized by the strongly ferromagnetic outer layers can mediate a significant amount of angular momentum over distances much exceeding the atomic spacing. The thickness of the spacer in our case is 6~nm, so no significant direct spin-transfer between the F-layers due to the conduction electrons is expected (RKKY essentially vanishes beyond 3~nm of Cr~\cite{PhysRevLett.65.2732,fert2,PhysRevLett.64.2304}). Therefore, in the model below, which uses a long-range interaction term, we restrict the action to the spacer spins only (40 monolayers for 6~nm thick spacer). The spins in the spacer are affected by this long-range exchange from both F and F$_{\text{p}}$ and, effectively, form a spin-chain mediating a relatively small but finite F$_\text{p}$-F exchange over several nanometers, at temperatures much exceeding the intrinsic Curie point of the spacer alloy where the material is nominally paramagnetic.

\begin{figure}[t]
\centering
\includegraphics[width=0.34\textwidth]{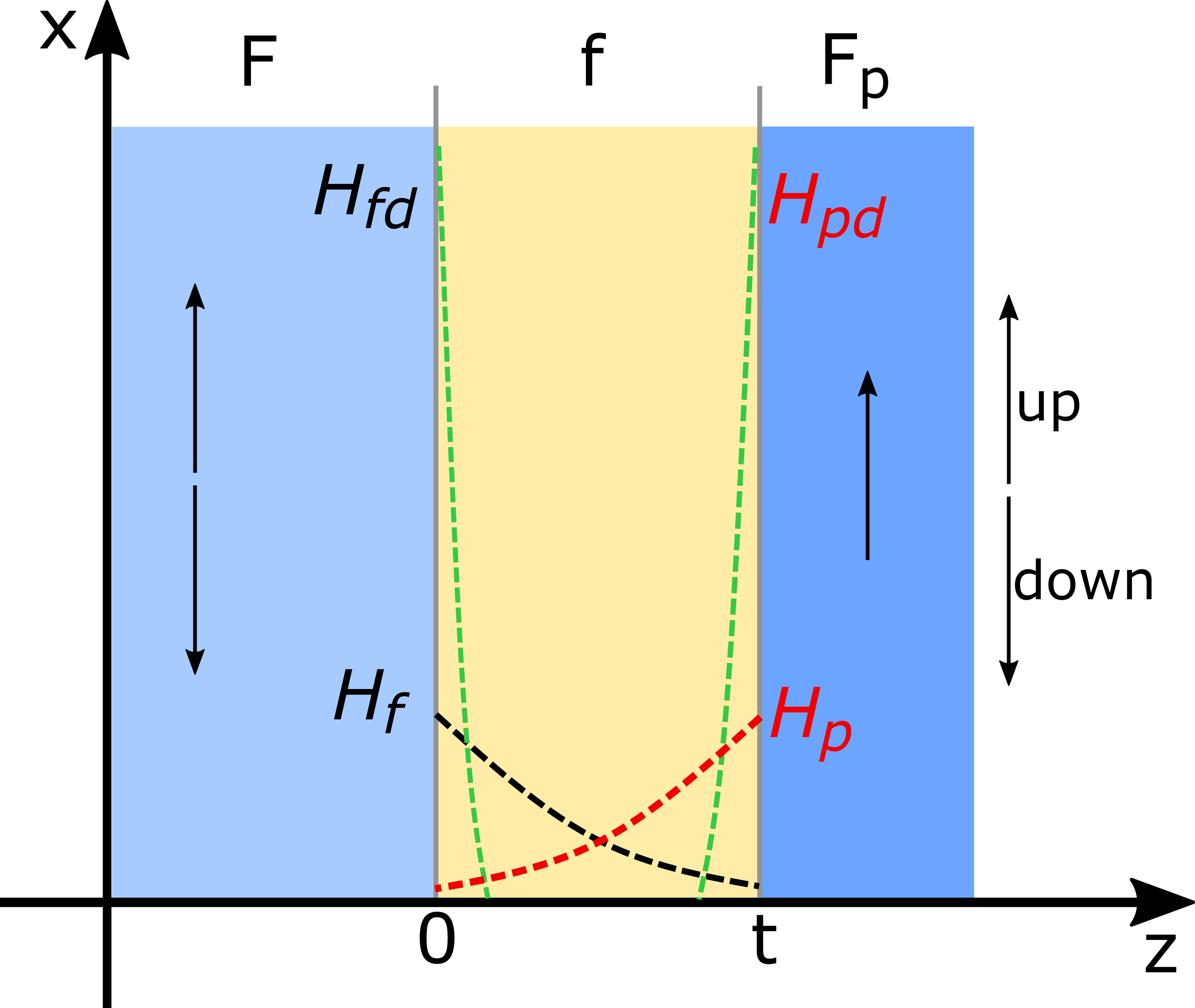}
\caption{\label{fig:profilesSchematic} Schematic of $\text{F}_\text{p}/$f$/\text{F}$ multilayer and profiles of effective exchange fields. Arrows within F and F$_\text{p}$ layers indicate up and down direction of layer's magnetization.}
\end{figure}

The need for a long-range spin-spin interaction term in quantitative modeling of the studied system can be illustrated with the the following simplified argument based on our MCE vs $T$ data of figure~\ref{fig:Tadcomparison}. For the spacer in a paramagnetic state (at $T>T_\text{C}$), its magnetic moment would be
\begin{equation}
\mathbf{M}_{\text{sp}}(z) = \chi \mathbf{H}_{\text{eff}}(z)\ ,
\label{eqn:spmag}
\end{equation}
where $\chi$ is the magnetic susceptibility of the paramagnetic spacer and $\mathbf{H}_{\text{eff}}$ is the effective field within the spacer, which depends on the $z$ coordinate normal to the plane of the $\text{F}_\text{p}/$f$/\text{F}$ stack (figure~\ref{fig:profilesSchematic}):
\begin{equation}
    \mathbf{H}_{\text{eff}}(z) = \mathbf{H}_{\text{ext}} + \mathbf{H}_{\text{f}}(z) + \mathbf{H}_{\text{p}}(z)+\mathbf{H}_{\text{fd}}(z,T) + \mathbf{H}_{\text{pd}}(z,T)\ .
\end{equation}
Here $\mathbf{H}_{\text{ext}}$ is external magnetic field, $\mathbf{H}_\text{f}(z)$ and $\mathbf{H}_\text{p}(z)$ are long-range effective field terms for the F/f and $\text{F}_\text{p}$/f interfaces, respectively, $\mathbf{H}_\text{fd}(z)$ and $\mathbf{H}_\text{pd}(z)$ are short-range effective exchange fields (direct-exchange proximity effect), which generally depend on both the temperature and coordinate, for the respective interfaces. At a temperature much above $T_\text{C}^\text{s}$, the magnetization induced in the paramagnetic spacer at its f/F interfaces by the direct exchange fields ($\mathbf{H}_\text{fd}$,$\mathbf{H}_\text{pd}$) is significant, however, the respective penetration depth (due to direct inter-atomic exchange) is much smaller than the total thickness of the spacer. As a result, the induced magnetization vanishes off the two interfaces toward the center of the spacer; their overlap is essentially zero and, therefore, the difference in magnetization between the parallel and antiparallel configurations of the F$_\text{p}$/f/F system expected from the short-range, direct exchange can be neglected. At temperatures comparable or lower than $T_\text{C}^\text{s}$, this short-range term does produce an overlap in magnetization (\emph{different} for P and AP, so $\Delta M \neq 0$ on P-to-AP switching) and must be taken into the account. A schematic of the multilayer with the spatial profile of the above effective fields is shown in figure~\ref{fig:profilesSchematic}. For the sake of this illustrative argument, we take the long-range exchange field (mediated by polarized electrons) as temperature independent, as it is expected to vary much weaker vs $T$ than the direct-exchange term, scaled by roughly the Fermi vs the Curie temperature.

Spin polarization of the electrons decays exponentially off the F/f interfaces into the spacer over a characteristic spin-flip length, $l_\text{sf}$, so the spatial dependence of the respective effective magnetic fields, $\textbf{H}_\text{f}(z)$ and $\textbf{H}_\text{p}(z)$, has the form 
\begin{equation}
\mathbf{H}_{\text{f}}(z) = \mathbf{M}_\text{f}J_\text{f}e^{-z/{l_{\text{sf}}}},\quad 
\mathbf{H}_{\text{p}}(z) = \mathbf{M}_\text{p}J_\text{p}e^{(z-t)/{l_{\text{sf}}}},\ 
\label{eqn:fieldProfile}
\end{equation}
where $t$ is the thickness of the spacer, $J_\text{f}$ and $J_\text{p}$ -- properly normalized exchange constants between the pairs of materials $\text{F}/$f and $\text{F}_\text{p}/$f, $\mathbf{M}_\text{f}$ and $\mathbf{M}_\text{p}$ -- magnetization vectors of the free and pinned layers, respectively. In figure~\ref{fig:profilesSchematic} the fields for the long-range exchange are shown out of proportion for visual clarity -- estimated to be about two orders of magnitude smaller than the corresponding short-range exchange fields (2\% range used in the numerical simulations below).

With the magnetization of the spacer denoted as $\textbf{M}_\text{sp}(z)$, the energy areal density is
\begin{equation}
\varepsilon(z) = \mathbf{M}_{\text{sp}}(z)\mathbf{H}_{\text{eff}}(z)\ .
\label{eqn:epsilon}
\end{equation}
Using \eqref{eqn:spmag} and integrating \eqref{eqn:epsilon} over $z$, the magnetic energy per volume of the paramagnetic spacer becomes
\begin{equation}
E = \chi\int_0^t\mathbf{H}_{\text{eff}}^2(z)\text{d}z\ .
\end{equation}
The energy difference, $\Delta E$, between the P ($\uparrow\uparrow$) and AP ($\uparrow\downarrow$) states of the trilayer is then
\begin{equation}
\Delta E = \chi\left[\int_0^t\mathbf{H}_{\text{eff}}^{\uparrow\uparrow 2}(z)\text{d}z - \int_0^t\mathbf{H}_{\text{eff}}^{\uparrow\downarrow 2}(z)\text{d}z\right].
\end{equation}

\begin{figure}[t]
\centering
\includegraphics[width=0.45\textwidth]{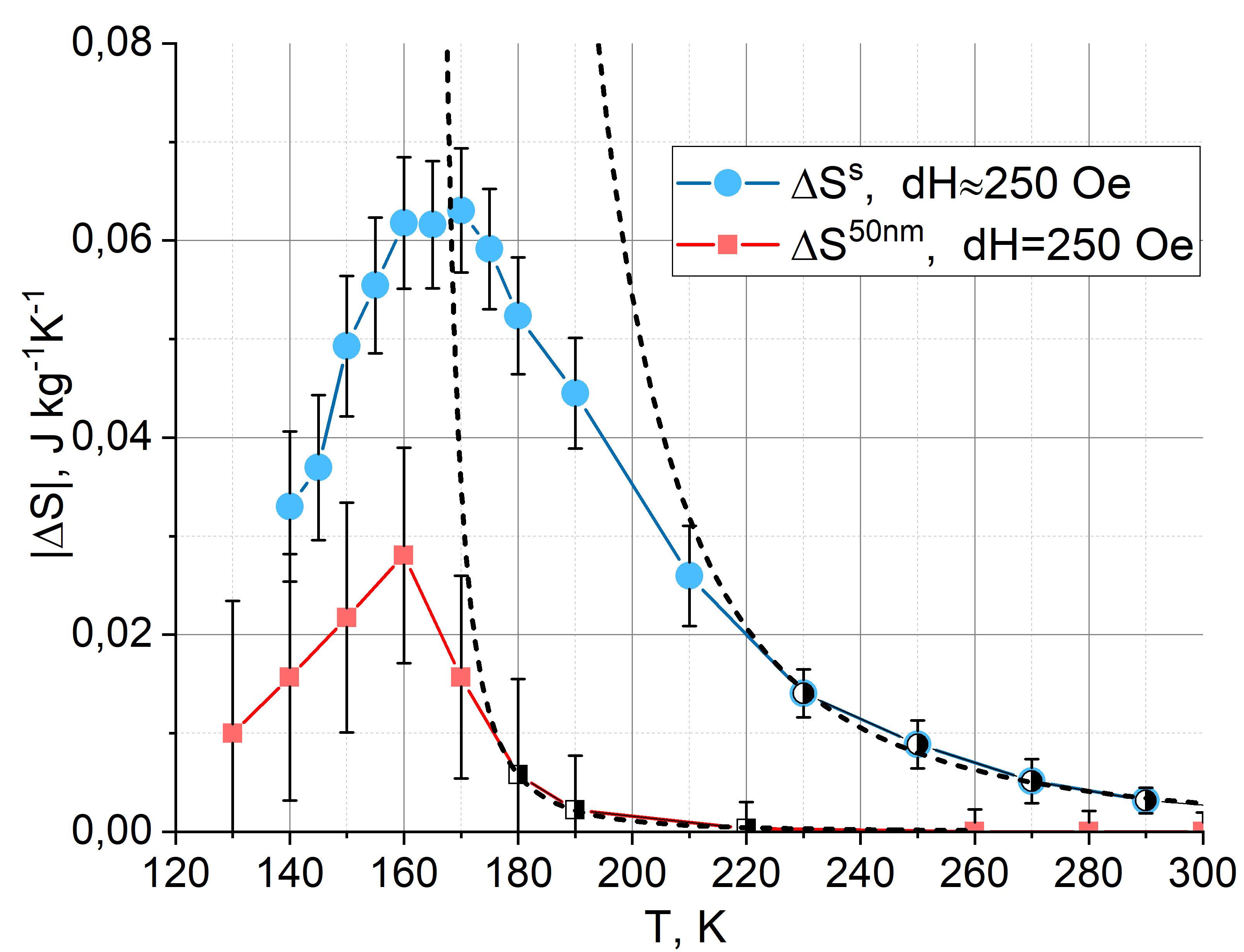}
\caption{\label{fig:DeltaSexperimental} Entropy change obtained from $\Delta T_{ad}$ data of figure~\ref{fig:Tadcomparison} for samples mA (trilayer; blue circles) and mC (bulk alloy; red squares); $\Delta H=250$~Oe. Half filled symbols mark data used for fitting $\Delta S(T)\propto\tau^{-k}$ at $T>T_\text{C}$ (dotted lines), with parameters found in main text.}
\end{figure}

Taking $H_\text{DC}$ as the field-offset and $H_\text{AC}$ as the width of the P-AP transition, the effective fields for the two states projected on to the P/AP axis ($x$) are
\begin{equation}
\begin{split}
H_{\text{eff}}^{\uparrow\uparrow}(z)=&H_\text{DC}+H_\text{AC}+ M_\text{f}J_\text{f}e^{{-z}/{l_{\text{sf}}}}+M_\text{p}J_\text{p}e^{{(z-t)}/{l_{\text{sf}}}},\\
H_{\text{eff}}^{\uparrow\downarrow}(z)=&H_\text{DC}-H_\text{AC}- M_\text{f}J_\text{f}e^{{-z}/{l_{\text{sf}}}}+M_\text{p}J_\text{p}e^{{(z-t)}/{l_{\text{sf}}}}.
\end{split}
\label{eqn:Heffs}
\end{equation}
According to the Landau theory of phase transitions, above the Curie point, $\chi\propto C\tau^{-\gamma}$, where $\tau=T-T_\text{C}$ and $\gamma$ is the critical exponent for the Curie transition. With constant heat capacity $C$, this gives for the entropy change on P-AP switching versus temperature
\begin{equation}
\Delta S = \frac{\partial\Delta E}{\partial\tau} = -C\gamma\tau^{-(\gamma + 1)}
\int_0^t\left [H_{\text{eff}}^{\uparrow\uparrow 2}(z) - H_{\text{eff}}^{\uparrow\downarrow 2}(z)\right ]\text{d}z\,
\label{eqn:deltaS}
\end{equation}
or $\Delta S(T)\propto\tau^{-k}$, with $k=\gamma+1$. figure~\ref{fig:DeltaSexperimental} shows $\Delta S$ converted from the experimentally measured $\Delta T_\text{ad}$ data of figure~\ref{fig:Tadcomparison}, alongside $C(T-T_{C})^{-k}$ fitting at high temperatures, with $T_\text{C}=162$~K and $k=2.28\pm0.25$ and $2.23\pm0.1$ for the mA and mC samples, respectively. The fitting yields the critical index, $\gamma$, close to 1.3 \emph{for both samples}, which is rather interesting as this value is characteristic for paramagnetic $\chi(T)$ and, at first glance, not expected to apply to the F/f/F structure. The explanation of the high-temperature critical-exponent behavior discussed below sheds additional light on the MCE mechanism at play in the studied system. 

For the bulk-like sample (mC), obviously, both short- and long-range effective exchange fields are absent as there are no interfaces, so only the external fields affect the material. In this case, at a given temperature \eqref{eqn:Heffs} and \eqref{eqn:deltaS} reduce  to $\Delta S(H)\propto H_\text{DC}H_\text{AC}$. With $H_\text{DC}=\text{const}$, $\Delta T_\text{ad}$ (or, equivalently, $\Delta S(H)$) should be linearly proportional to $H_\text{AC}$, which is indeed observed experimentally; see figure~\ref{fig:slopes}, red triangles for sample mC.

In contrast, for the F$_\text{p}$/f/F trilayer, the P-AP switching of the magnetization results in a rather sharp, step-like behavior, with a characteristic saturation at a field where the magnetization of the free layer is fully rotated; figure~\ref{fig:slopes}, blue circles for sample mA. The effect is normalized to the volume of the MCE active layer (spacer or thick film), so the fact that the initial $\Delta T_\text{ad}$ vs $H_\text{AC}$ of the trilayer is much steeper than that of the thick film indicates that the effective (exchange) field producing the response is much stronger than the externally applied field. Naturally, as the magnetization vector of the free layer rotates, its proximity field within the spacer (mostly long-range component at higher temperatures and mostly direct exchange near $T_\text{C}$) follows the rotation and partially cancels out that from the pinned layer when the rotation is complete (AP state). The resulting relatively large (as it is of exchange origin) effective-field differential as given by \eqref{eqn:deltaS} is expected to scale with the angle of rotation, produce a steep slope, and saturate when in the AP state. This picture applies above as well as near the Curie point and describes well the experimentally observed behavior, shown in figure~\ref{fig:slopes}. 

Interestingly, the fact that the 295~K data has the same functional form of step-saturation as the 170~K data (figure~\ref{fig:slopes}, green diamonds), justifies the assumption that a weakly temperature dependent, long-range exchange is present in the structure at temperatures much higher than the transition point. Otherwise, with the direct-exchange proximity fields (very short-range at high-$T$) fast-decaying and not overlapping in this regime, the external field would be expected to produce a linear slope of the type seen in the mC sample data. In the Simulations section below, we will discuss in detail domain-wall like spin distributions within the spacer in the AP state responsible for this MCE behavior and how they vary with temperature and exchange characteristics.

The above qualitative argument offer a compelling illustration of the mechanisms behind the studied phenomenon. However, the presence of the internal exchange inside the spacer material, including the range above its Curie temperature, should be expected to contribute and likely modify, at least quantitatively and potentially significantly, the MCE in the F/f/F system near and below $T_\text{C}^\text{s}$. We therefore have conducted in-depth numerical simulations of the system in a wide parameter space, including direct-proximity as well as long-range exchange interactions, for which a computationally efficient phenomenological model  has been developed for large phase-space mapping of our system, which is difficult to do using other approaches, such as the well-established but more resource intensive atomistic spin dynamics (discussed in detail elsewhere~\cite{Milton_2022}; showing good qualitative agreement with the conclusion herein in regions of overlapping parameter space, e.g., only short-range exchange). The results of the numerical simulations are presented below and show that the AP state of the trilayer near and above $T_\text{C}$ has a nontrivial spin configuration similar to a Bloch domain wall, and it is this configuration left out by the paramagnetic-spacer model that actually determines the strength of the MCE in the system and its variation with temperature and composition.

\subsection{\label{ssec:numerical}Numerical Simulations}
\begin{figure*}[t]
\centerline{

\begin{tikzpicture}[font=\sffamily]
\begin{scope}
\node (a) at (0,0) {\includegraphics[scale=0.28]{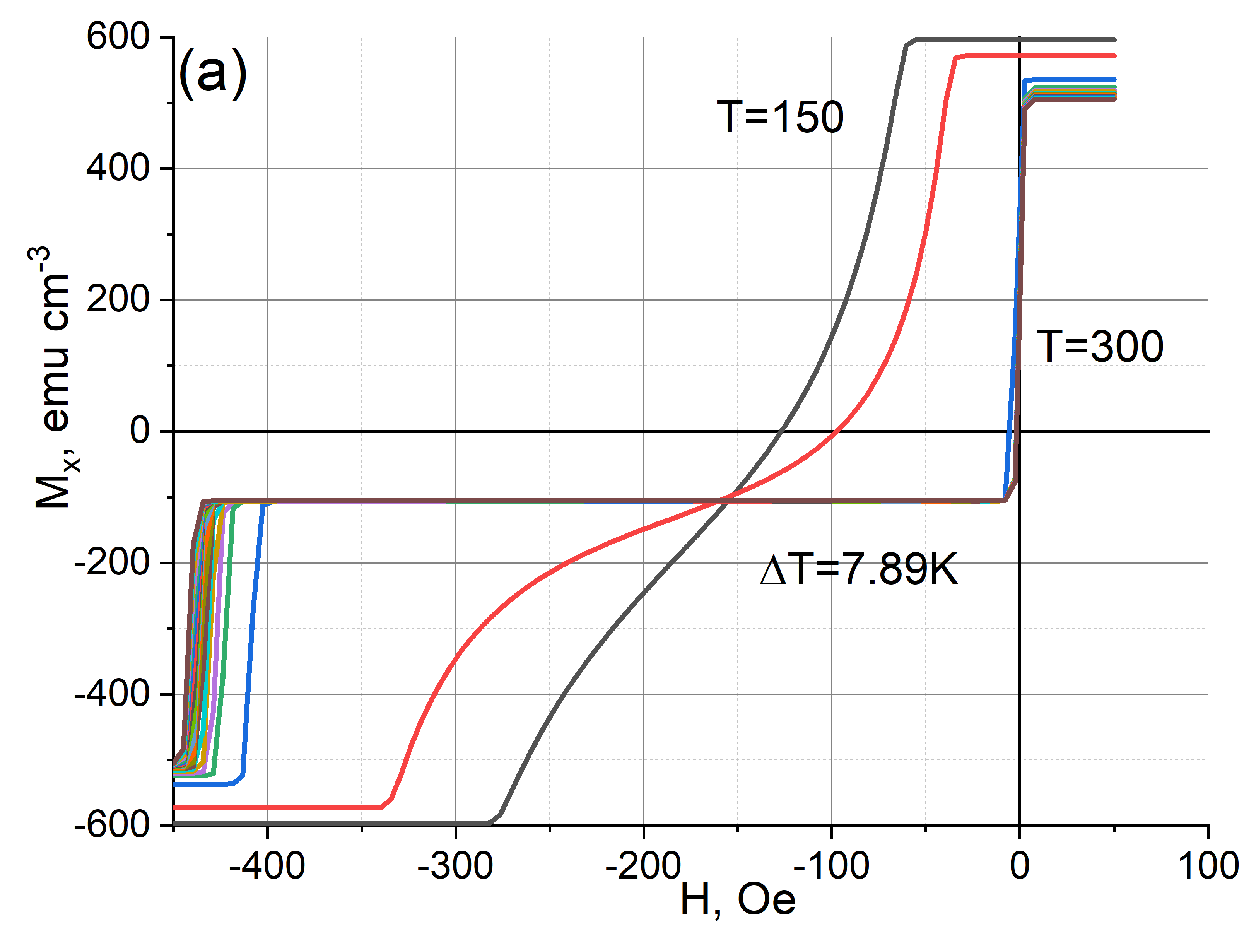}};
\node (b) at (8,0) {\includegraphics[scale=0.28]{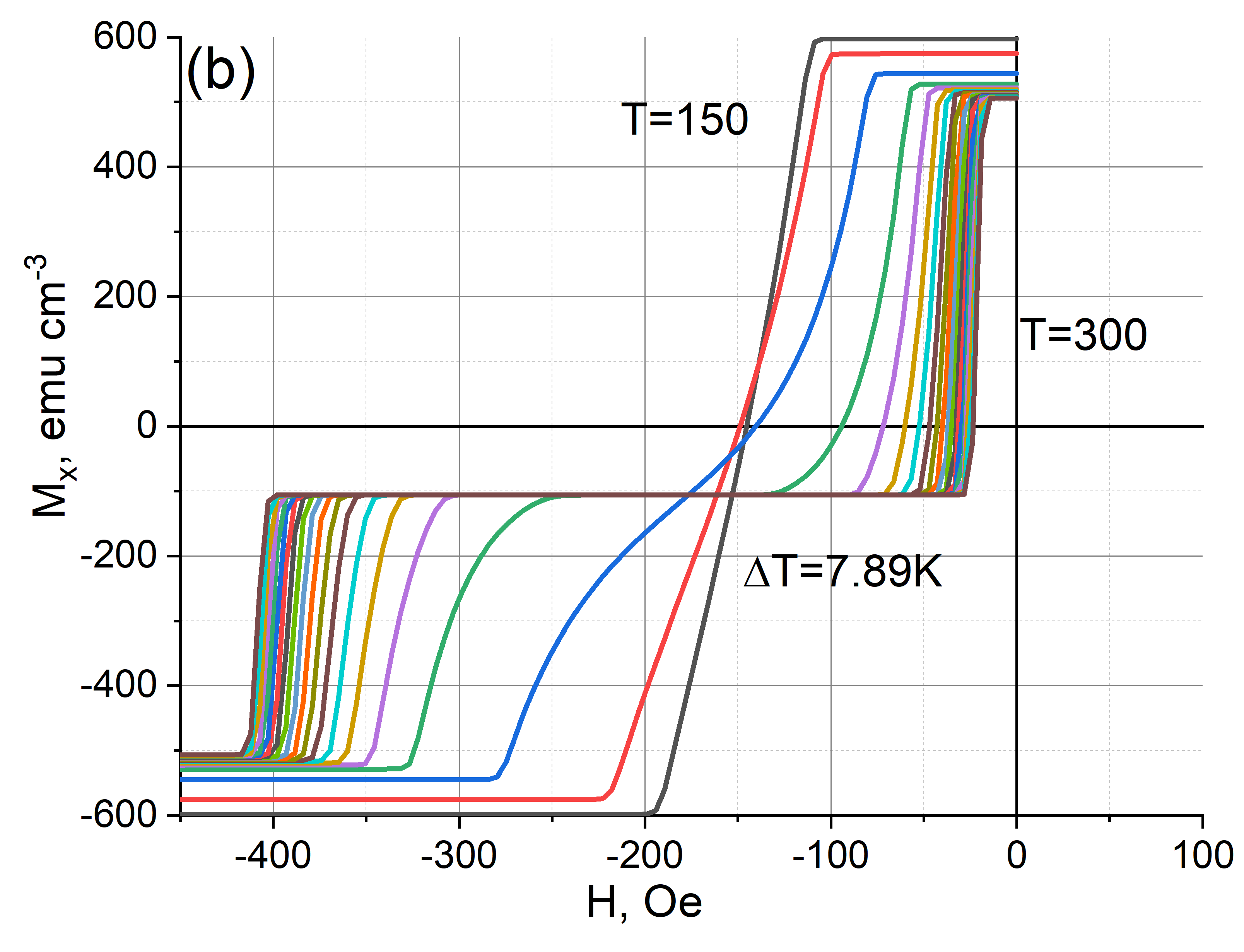}};
\end{scope}

\end{tikzpicture}
}
\caption{\label{fig:simComparison}Magnetization reversal loops of $\text{F}_\text{p}$/f/F system simulated to fit measured $M-H$ loops of figure~\ref{fig:4loops}: (a) without long-range exchange, with short-range exchange only; (b) with long-range exchange as well as short-range exchange.}
\end{figure*}

Atomistic spin simulations of our system~\cite{Milton_2022}, which would incorporate long-range spin-spin interactions (several nanometers, full width of spacer, upto 40 nearest neighbors), would be extremely time and resource consuming. We, therefore, develop here a phenomenological computational model, simulating the F/f/F system as a stack of magnetization vectors representing atomic monolayers throughout the trilayer. The magnitude of the magnetization of a monolayer is defined by the Brillouin function of temperature and the local effective field, consisting of a Zeeman term and the intra- and inter-monolayer exchange contributions. The direction of the magnetization vector is the same as the direction of the local effective field acting on it. The strong out-of-plane demagnetization in our thin film geometry restricts the spins orientationally to the film plane. One of the ferromagnetic layers is `pinned' by a fixed static field of 450 Oe, the other has no in-plane anisotropy and would be free to rotate in a variable external field in case the spacer was fully nonmagnetic.

The overall minimization algorithm consists of separate consecutive steps of recursive calculations of the magnetization ($M$) and a minimization of its orientation ($\varphi$) by the gradient descent method -- this sequence constitutes a single `iteration'. The exit from the iteration loop is possible on two conditions - either $\Delta M<10^{-7}$ $\text{emu cm}^{-3}$ and $\Delta \varphi <10^{-6}$ deg are reached, or the maximum number of iterations is reached. The final state of the system for various parameter combinations showed only minor changes after about 90k iterations, which was taken to be the maximum simulation length, sufficient to achieve good accuracy.

The values of the saturation magnetization and Curie temperature used in the simulations were obtained from the literature~\cite{crangle1971magnetization}. The Curie temperature of the spacer was chosen to be 160~K as this value was observed on the experiment. The Curie temperature of the free and pinned ferromagnetic layers was chosen to be 2000~K to minimize the variation of the F-layers' magnetization with temperature. This approximation speeds up the simulations and essentially limits the variable part of the system to only the spacer, as we specifically aim to investigate the effect of the properties of the spacer on the overall behaviour of the trilayer. The strength of the intrinsic exchange in the spacer was chosen so as to reproduce the experimentally observed Curie temperature (160~K). The strength of the short-range proximity exchange between the FeCr and Py layers (at F/f interfaces) generally depends on the quality of the interfaces and, for our samples, was numerically estimated to be about 150\% of the intrinsic exchange in the spacer.

To qualitatively reproduce the magneto-statics observed on the experiment, two simulations of the F/f/F system were performed using the aforementioned algorithm. We have compared the magnetostatic behaviour of the system for the case where the only exchange interaction available is direct exchange with the nearest and the next nearest neighbours, and the case where the direct short-range exchange is supplemented by long-range exchange interactions between the interface spins of the F layers and the spins in all of the spacer (with spatially decaying strength). In what follows, we refer to these two types of simulations as ``without long-range exchange" and ``with long-range exchange". The long-range exchange constant is taken to have the same $z$-dependence as the corresponding exchange field given by \eqref{eqn:fieldProfile}:
\begin{equation}
J_\text{f}(z) = J_\text{0f}e^{{-z}/{l_{\text{sf}}}},\quad 
J_\text{p}(z) = J_\text{0p}e^{{(z-t)}/{l_{\text{sf}}}},\ 
\label{eqn:ExchangeProfile}
\end{equation}
where $J_\text{f}(z)$ and $J_\text{p}(z)$ are the $z$-dependent exchange constants proximity-induced within the spacer by the free and pinned F/f interfaces, $J_\text{0f}$ and $J_\text{0p}$ -- their interface values (at $z=0$), $z$ -- the coordinate normal to the film plane, $l_\text{sf}$ - the characteristic decay length of the long-range exchange interaction (the spin-flip length in the spacer, assuming the effect is due to polarized electrons).

The simulated $M-H$ loops for the two model cases, namely, without and with long-range exchange, are shown in figure~\ref{fig:simComparison}. Parameters $l_\text{sf}=4$~nm and $J_\text{0f}=J_\text{0p}=$ 2\% of the intrinsic exchange in the spacer were found to optimally fit the measured data of figure~\ref{fig:4loops} (although likely not the only one possible, this parameter combination is representative of the optimal range and is quite reasonable based on the general considerations the energetics and scales of spin relaxation and spin transfer in dilute ferromagnetic alloys). It is clearly visible that with only the direct exchange, the coupling between the $\text{F}_\text{p}$ and $\text{F}$ layers through a 6-nm spacer fully vanishes at $T>T_\text{C}^\text{s}$, as the minor loop of the F layer is centered at zero field. In contrast, for the case with both the direct and long-range exchange, the inter-layer coupling survives up to room temperature, yielding a nonzero exchange bias of the free layer. This specific feature of the simulated magnetization reversal loop, which we find to be critically dependent on the presence of long-range exchange, reproduces the behavior observed experimentally in (figure~\ref{fig:4loops}(b-d)) and previously in~\cite{Vdovichev2018High,Lim2013Temperature-dependent,Kravets2012Temperature-controlled,Kravets2014Synthetic,Magnus2016Long-range}. 

\begin{figure*}[t!]
\centerline{
\begin{tikzpicture}[font=\sffamily]
\begin{scope}
\node (a) at (0,0) {\includegraphics[scale=0.28]{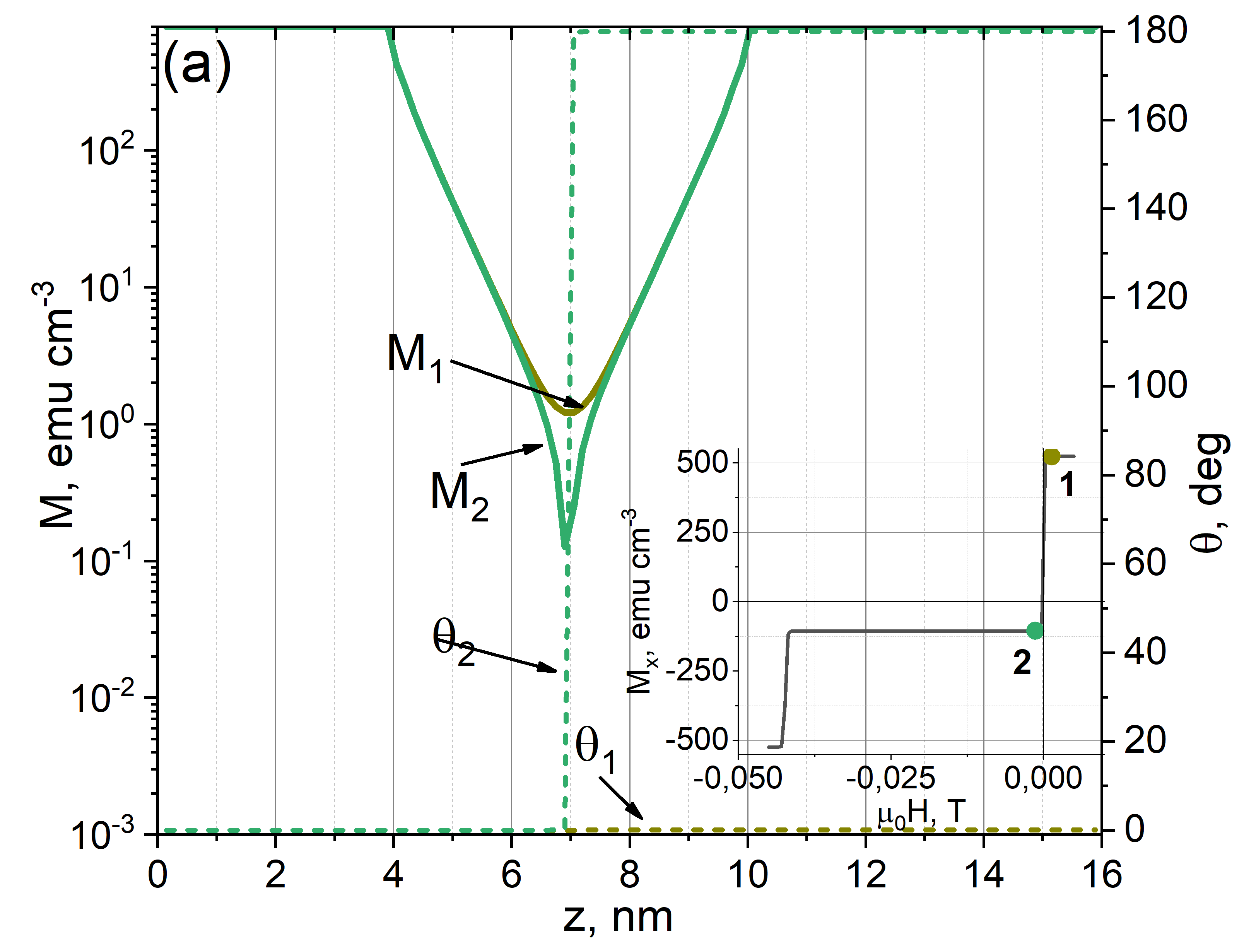}};
\node (b) at (8,0) {\includegraphics[scale=0.28]{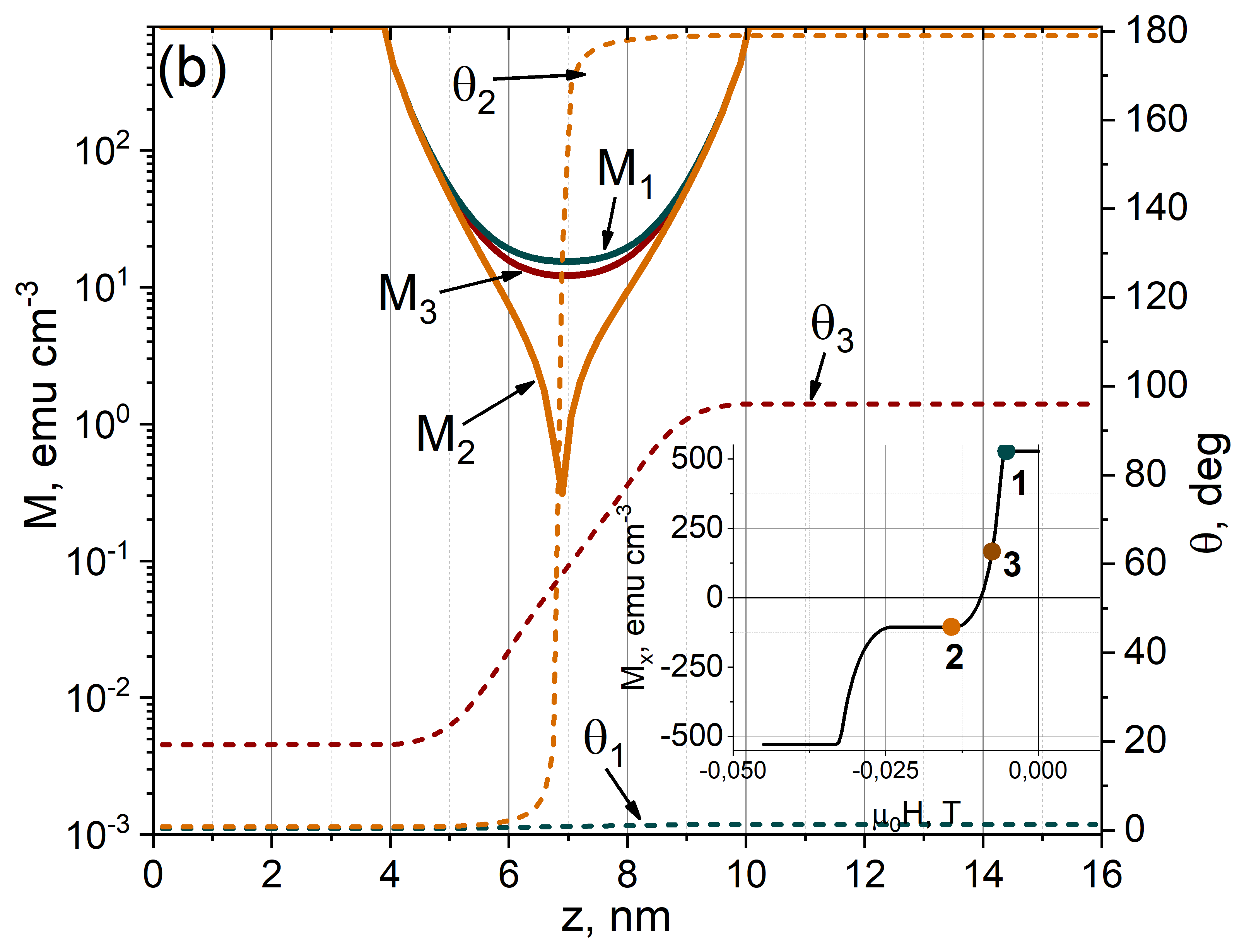}};
\end{scope}
\end{tikzpicture}
}
\caption{\label{fig:profiles}Simulated magnetization magnitude and orientation through thickness ($z$) of F$_\text{p}$/f/F system: (a) without and (b) with long-range exchange; $z=0\div 4$~nm is occupied by pinned layer, $4\div 10$~nm -- spacer, $10\div 16$~nm -- free layer. Solid lines indicate magnitude of magnetization vector (left axis scale) and dashed lines indicate its orientation (right axis scale). Insets: magnetization reversal loops with characteristic points indicated by numerals, $n=1,2,3$, where point $n$ corresponds to magnetization profile $M_n(z)$ and orientation profile $\theta_n(z)$. $T=174$~K.}
\end{figure*}

A completed iteration sequence yields the equilibrium magnitude and orientation of the magnetization of each monolayer, for a given parameter set. As an example, the evolution of the simulated magnetization profile during the switching of the free layer, with the long-range exchange disabled and enabled, are shown in figure~\ref{fig:profiles}(a) and (b), respectively, for $T$=174~K. Analyzing and comparing the data in figure~\ref{fig:profiles} (a) and (b), several important observations can be made. In the parallel state and with only short-range exchange enabled (point 1 in (a)), the proximity-induced magnetization decays exponentially from the F/f interfaces toward the center of the spacer, which in fact is expected for penetration of the magnetic order parameter in the Landau theory (for individual interfaces). Despite the fact that the direct exchange may induce substantial magnetization inside the spacer in the direct proximity of the F layers, this magnetization decays rapidly and vanishes toward the center of the spacer, and therefore can not provide either a significant change in magnetization on P-to-AP switching or any appreciable coupling between the outer F layers. With both direct and long-range exchange present (figure~\ref{fig:profiles}(b)), the P-state magnetization deviates from the above behavior and is an order of magnitude higher in the middle of the spacer compared to that for only direct exchange. A similar tendency is observed for the AP-state (point 2 in (a) and (b)) - the magnetization profile functionally deviates from an exponential decay and this deviation is much stronger in the case where the long-range exchange is enabled. Due to a cancellation of the opposing effective exchange fields (AP-state), overlapping in the center of the spacer, the magnetization at $z=7$~nm is drastically decreased, with the difference on P-to-AP switching being much bigger when the long-range exchange is present. The intermediate-state orientational profile (figure~\ref{fig:profiles}(b), point 3, dashed line for $\theta_3 (z)$) reveals a scissor-like state in the system, with a $\sim 90$-degree `exchange spring' in the spacer, which must be the reason for the absence of any significant cancellation of the effective exchange at $z=7$~nm and hence no significant additional magnetic disorder. In contrast, the orientational transition is sharp in the AP-state ($\theta_2 (z)$) with a well defined 180-wall, a single nanometer in thickness, producing strong local magnetic disorder.

The switching of the free layer in the model with only direct exchange (inset to figure~\ref{fig:profiles}(a)) is step-like and no intermediate state is observed between the parallel and antiparallel configurations at all temperatures higher than the Curie temperature. Enabling long-range exchange results in non-vanishing interlayer coupling in a wide temperature range above $T_\text{C}^\text{s}$ and, consequently, in rotational states intermediate between P and AP. Due to this coupling, both magnetization vectors of the pinned and free layers tilt to an equilibrium orientation that is generally different from the direction of the external field (e.g., point 3 in inset to figure~\ref{fig:profiles}(b)). The magnetization magnitude in the spacer decreases only slightly from its value in the P state for intermediate angles. Only in the vicinity of the AP state the spacer demagnetizes significantly in the very center, showing a narrow orientational spin transition similar to a Bloch domain wall. 

\begin{figure}[t]
\centering
\includegraphics[width=0.45\textwidth]{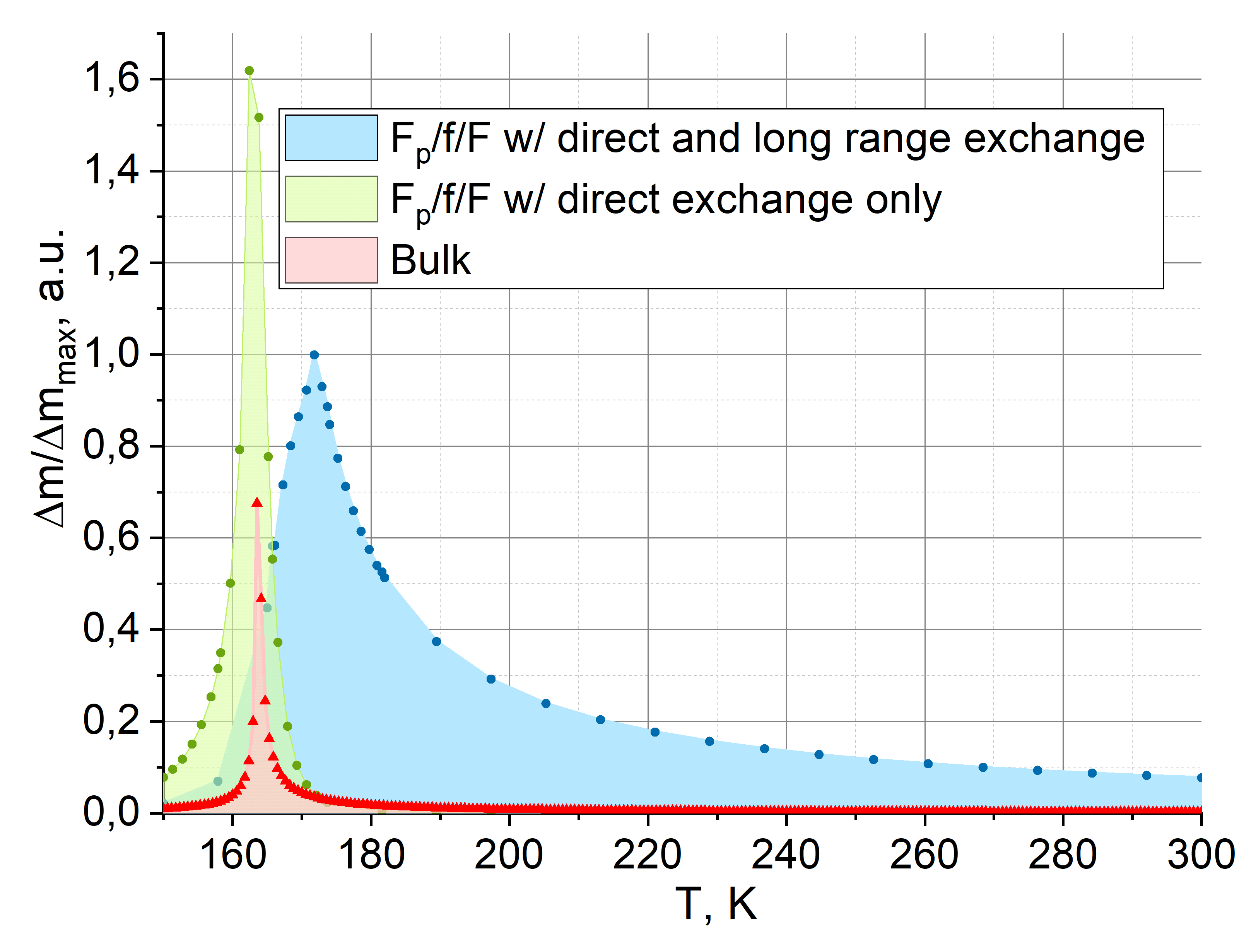}
\caption{\label{fig:DeltaMsimulated} Simulated temperature dependence of magnetic disorder $\Delta m$ induced by P-to-AP switching for trilayer with \emph{both} direct and long-range exchange (blue) and \emph{only} direct exchange (green), in units of former. Simulated $\Delta m$ for bulk Fe-Cr alloy (thick film) in applied field of 250~Oe (corresponds to switching field of trilayer), likewise normalized (red).}
\end{figure}

The numerical simulations have a high illustrative and interpretational value as they yield the spatial spin profiles in the structure, which are difficult, if at all possible, to obtain on the experiment -- VSM magnetometry yields only a projection of the net magnetic moment of the structure as a whole. Having the microscopic spin distribution sheds light on the mechanism behind the system's magneto-caloric response and its strength. Thus, extracting the change in the magnetization magnitude on P-to-AP switching yields the change in magnetic disorder (or entropy of the magnon sub-system; between points 1 and 2 in the inset to figure~\ref{fig:profiles}(a) and (b)). This normalized magnetization difference is shown in figure~\ref{fig:DeltaMsimulated}, simulated with (blue) and without (green) long-range exchange. For comparison, the magnetization change in the bulk spacer material in response to 250 Oe of external field is shown in red. To convert the magnetization change to the magnetic entropy change, different approaches can be used, but the general relation is that $\Delta S \propto \Delta m$. 

The key result presented in figure~\ref{fig:DeltaMsimulated} is that for the trilayer system modelled with the long-range exchange enabled (blue) the strong magnetic disorder (MCE) induced by the P-AP transition persists to temperatures significantly higher than $T_\text{C}^s$, with the maximum $\Delta m$ shifted to slightly higher than $T_\text{C}^\text{s}$. This is in excellent agreement with the results of our direct MCE measurements shown in Figs.~\ref{fig:Tadcomparison},\ref{fig:DeltaSexperimental}. At the same time, $\Delta m$ for the case of the trilayer with no long-range exchange as well as for the bulk system (simulated with direct exchange only) peak at $T_\text{C}^\text{s}$ and vanish directly above it. It is clear that the strength of the field-induced demagnetization is enhanced in the case of the F/f/F trilayer system, both with and without long-range exchange taken into account, compared to the case of the bulk alloy (thick film). Estimating potential MCE efficiency of the proximity-enhanced material versus that of the bulk spacer alloy, we note that the ratio of areas under $\Delta m$ of the respective materials is proportional to the ratio of their relative cooling power (RCP)~\cite{ELHAFIDI2018500}, which means that the RCP of the proximity-enhanced spacer (simulated with long-range exchange) can be up to two order of magnitude higher than that of its constituent material. The simulated case with only direct exchange (green in figure~\ref{fig:DeltaMsimulated}) does show enhanced MCE and RCP, however it clearly disagrees with the experiment as to the peak position and, especially, the high temperature range where the effect is pronounced. 

Each magnon corresponds to one Bohr magneton of thermal demagnetization (one $\hbar$ in angular momentum), distributed in energy up to $k_B T_\text{C}^\text{s}$. Taking as an estimate the mean magnon energy for our spacer as 10 meV and the simulated peak value of $\Delta m=0.6$~emu/cm$^{-3}$ for the trilayer, we arrive at the peak adiabatic temperature change of approximately 30~mK, which agrees well with 25~mK obtained on the experiment (figure~\ref{fig:Tadcomparison}).

Comparing the simulated and the experimental results (figure~\ref{fig:DeltaMsimulated} versus figure~\ref{fig:DeltaSexperimental}) we conclude that long-range exchange is necessary to adequately explain and quantify the observed MCE effect. The long-range exchange affects the magnetization profile within the spacer in a significant way, magnetizing the otherwise paramagnetic material into a domain-like spin distribution of spatially variable magnitude, which shows a strong change on P-to-AP switching in the structure, in a comparatively very low external field ($\sim 100$~Oe, $2\div 3$ orders of magnitude lower than for conventional ADR systems). Long-range exchange is the key to have MCE peak noticeably above $T_\text{C}^\text{s}$ and extend in a strong fashion to temperatures significantly higher than $T_\text{C}^\text{s}$, which should be useful for extending the material's operating range in device applications. Our experiment and simulations confirm that the concept of enhanced MCE in F/f/F systems is valid both in terms of higher peak entropy-change values as well as much higher RCP.

\section{\label{sec:conclusion}Conclusion}
We have performed, using a membrane-based nano-calorimetry method, direct measurements of the adiabatic temperature change in magnetic multilayers designed for enhanced MCE. The measurements show that the volume and field normalized MCE induced by the magnetic proximity effect in a specially designed trilayer can be significantly higher than the conventional MCE in the corresponding bulk material. The respective relative cooling power is vastly stronger, one to two orders of magnitude, due in particular to a much broader operating temperature range of the exchange-enhanced material. We therefore conclude that the approach of exchange-enhancing MCE in magnetic nanostructures is valid and effective in reducing the requirements on the externally applied magnetic field. The experimental MCE measurement technique was the key to these findings as the commonly used indirect magnetometry-based methods were found to be inaccurate when applied to the orientational magnetic transition in the studied multilayered system.

The obvious challenge with thin F/f/F type multilayered structures is the relatively small volume of the active material with enhanced MCE properties. This issue may be overcome by using spacer materials having intrinsically higher MCE as well as by vertical stacking of the core trilayer structure. The presence of the assisting F-layers roughly doubles the volume of the stack at no direct benefit to MCE, however, the fact that they can switch in low fields ($\sim$10~mT) at very high speed (sub-$\mu$s) make feasible high-frequency cooling cycles, orders of magnitude faster than the typical paramagnetic ADR cycles where ramping up/down $\sim 5$~T field over minutes is required. Particularly attractive may be cooling of micro/nano-devices (compact sensors, sources, gates, etc.) based on the very membrane platform we demonstrate, provided a mechanical arrangement for heat outflow can be integrated and synced with the AC field cycle.

We find that a long-range exchange interaction, alongside the direct exchange, must be taken into account in order to adequately describe the observed properties of the studied metallic trilayer system. This interaction can be due to the conduction electrons polarized by the strongly ferromagnetic outer layers, traversing the weakly ferromagnetic spacer, thereby mediating spin-spin exchange over several nanometers. Our phenomenological model, which includes such long-range exchange, has allowed efficient, wide parameter-space numerical simulations, which explain the experimentally observed behavior in an illustrative as well as quantitatively accurate way. The enhanced MCE effect is shown to be due to an exchange-spring in the spacer, pinched-off in the center on rotation into the antiparallel state of the device.

\begin{acknowledgments}
The authors are thankful to V.M. Kalita and Yu.I. Dzhezherya  for fruitful discussions. Support by Olle Engkvist Foundation (207-0460) and Swedish Research Council (VR 2018-03526) are gratefully acknowledged.
\end{acknowledgments}

\nocite{*}
\bibliography{ref}

\begin{thebibliography}{42}%
\makeatletter
\providecommand \@ifxundefined [1]{%
 \@ifx{#1\undefined}
}%
\providecommand \@ifnum [1]{%
 \ifnum #1\expandafter \@firstoftwo
 \else \expandafter \@secondoftwo
 \fi
}%
\providecommand \@ifx [1]{%
 \ifx #1\expandafter \@firstoftwo
 \else \expandafter \@secondoftwo
 \fi
}%
\providecommand \natexlab [1]{#1}%
\providecommand \enquote  [1]{``#1''}%
\providecommand \bibnamefont  [1]{#1}%
\providecommand \bibfnamefont [1]{#1}%
\providecommand \citenamefont [1]{#1}%
\providecommand \href@noop [0]{\@secondoftwo}%
\providecommand \href [0]{\begingroup \@sanitize@url \@href}%
\providecommand \@href[1]{\@@startlink{#1}\@@href}%
\providecommand \@@href[1]{\endgroup#1\@@endlink}%
\providecommand \@sanitize@url [0]{\catcode `\\12\catcode `\$12\catcode
  `\&12\catcode `\#12\catcode `\^12\catcode `\_12\catcode `\%12\relax}%
\providecommand \@@startlink[1]{}%
\providecommand \@@endlink[0]{}%
\providecommand \url  [0]{\begingroup\@sanitize@url \@url }%
\providecommand \@url [1]{\endgroup\@href {#1}{\urlprefix }}%
\providecommand \urlprefix  [0]{URL }%
\providecommand \Eprint [0]{\href }%
\providecommand \doibase [0]{https://doi.org/}%
\providecommand \selectlanguage [0]{\@gobble}%
\providecommand \bibinfo  [0]{\@secondoftwo}%
\providecommand \bibfield  [0]{\@secondoftwo}%
\providecommand \translation [1]{[#1]}%
\providecommand \BibitemOpen [0]{}%
\providecommand \bibitemStop [0]{}%
\providecommand \bibitemNoStop [0]{.\EOS\space}%
\providecommand \EOS [0]{\spacefactor3000\relax}%
\providecommand \BibitemShut  [1]{\csname bibitem#1\endcsname}%
\let\auto@bib@innerbib\@empty
\bibitem [{\citenamefont {Pecharsky}\ and\ \citenamefont
  {Gschneidner}(1997)}]{Pecharsky1997Giant}%
  \BibitemOpen
  \bibfield  {author} {\bibinfo {author} {\bibfnamefont {V.~K.}\ \bibnamefont
  {Pecharsky}}\ and\ \bibinfo {author} {\bibfnamefont {K.~A.}\ \bibnamefont
  {Gschneidner}, \bibfnamefont {Jr.}},\ }\bibfield  {title} {\bibinfo {title}
  {Giant magnetocaloric effect in $\text{Gd}_5(\text{Si}_2\text{Ge}_2$)},\
  }\href {https://doi.org/10.1103/PhysRevLett.78.4494} {\bibfield  {journal}
  {\bibinfo  {journal} {Physical Review Letters}\ }\textbf {\bibinfo {volume}
  {78}},\ \bibinfo {pages} {4494} (\bibinfo {year} {1997})}\BibitemShut
  {NoStop}%
\bibitem [{\citenamefont {Yu}\ \emph {et~al.}(2003)\citenamefont {Yu},
  \citenamefont {Gao}, \citenamefont {Zhang}, \citenamefont {Meng},\ and\
  \citenamefont {Chen}}]{Yu2003Review}%
  \BibitemOpen
  \bibfield  {author} {\bibinfo {author} {\bibfnamefont {B.}~\bibnamefont
  {Yu}}, \bibinfo {author} {\bibfnamefont {Q.}~\bibnamefont {Gao}}, \bibinfo
  {author} {\bibfnamefont {B.}~\bibnamefont {Zhang}}, \bibinfo {author}
  {\bibfnamefont {X.}~\bibnamefont {Meng}},\ and\ \bibinfo {author}
  {\bibfnamefont {Z.}~\bibnamefont {Chen}},\ }\bibfield  {title} {\bibinfo
  {title} {Review on research of room temperature magnetic refrigeration},\
  }\href {https://doi.org/10.1016/S0140-7007(03)00048-3} {\bibfield  {journal}
  {\bibinfo  {journal} {International Journal of Refrigeration}\ }\textbf
  {\bibinfo {volume} {26}},\ \bibinfo {pages} {622} (\bibinfo {year}
  {2003})}\BibitemShut {NoStop}%
\bibitem [{\citenamefont {Metz}\ \emph {et~al.}(2005)\citenamefont {Metz},
  \citenamefont {Kuijpers}, \citenamefont {Solomon}, \citenamefont {Andersen},
  \citenamefont {Davidson}, \citenamefont {Meyer}, \citenamefont {on~Climate
  Change. Response Strategies Working~Group}, \citenamefont {on~Climate Change.
  Working Group~I},\ and\ \citenamefont {{UNEP. Technology and Economic
  Assessment Panel}}}]{Metz_2005}%
  \BibitemOpen
  \bibfield  {author} {\bibinfo {author} {\bibfnamefont {B.}~\bibnamefont
  {Metz}}, \bibinfo {author} {\bibfnamefont {L.}~\bibnamefont {Kuijpers}},
  \bibinfo {author} {\bibfnamefont {S.}~\bibnamefont {Solomon}}, \bibinfo
  {author} {\bibfnamefont {S.~O.}\ \bibnamefont {Andersen}}, \bibinfo {author}
  {\bibfnamefont {O.~R.}\ \bibnamefont {Davidson}}, \bibinfo {author}
  {\bibfnamefont {L.}~\bibnamefont {Meyer}}, \bibinfo {author} {\bibfnamefont
  {I.~P.}\ \bibnamefont {on~Climate Change. Response Strategies
  Working~Group}}, \bibinfo {author} {\bibfnamefont {I.~P.}\ \bibnamefont
  {on~Climate Change. Working Group~I}},\ and\ \bibinfo {author} {\bibnamefont
  {{UNEP. Technology and Economic Assessment Panel}}},\ }\href
  {https://digitallibrary.un.org/record/3892954?ln=en} {\emph {\bibinfo {title}
  {Safeguarding the ozone layer and the global climate system}}},\ edited by\
  \bibinfo {editor} {\bibfnamefont {I.~P.}\ \bibnamefont {on~Climate~Change}}\
  (\bibinfo  {publisher} {Cambridge University Press},\ \bibinfo {year}
  {2005})\BibitemShut {NoStop}%
\bibitem [{\citenamefont {Miller}\ \emph {et~al.}(2014)\citenamefont {Miller},
  \citenamefont {Belyea},\ and\ \citenamefont
  {Kirby}}]{Miller2014Magnetocaloric}%
  \BibitemOpen
  \bibfield  {author} {\bibinfo {author} {\bibfnamefont {C.~W.}\ \bibnamefont
  {Miller}}, \bibinfo {author} {\bibfnamefont {D.~D.}\ \bibnamefont {Belyea}},\
  and\ \bibinfo {author} {\bibfnamefont {B.~J.}\ \bibnamefont {Kirby}},\
  }\bibfield  {title} {\bibinfo {title} {Magnetocaloric effect in nanoscale
  thin films and heterostructures},\ }\href {https://doi.org/10.1116/1.4882858}
  {\bibfield  {journal} {\bibinfo  {journal} {Journal of Vacuum Science \&
  Technology A: Vacuum, Surfaces, and Films}\ }\textbf {\bibinfo {volume}
  {32}},\ \bibinfo {pages} {040802} (\bibinfo {year} {2014})}\BibitemShut
  {NoStop}%
\bibitem [{\citenamefont {Doblas}\ \emph {et~al.}(2017)\citenamefont {Doblas},
  \citenamefont {Moreno-Ramírez}, \citenamefont {Franco}, \citenamefont
  {Conde}, \citenamefont {Svalov},\ and\ \citenamefont
  {Kurlyandskaya}}]{Doblas2017Nanostructuring}%
  \BibitemOpen
  \bibfield  {author} {\bibinfo {author} {\bibfnamefont {D.}~\bibnamefont
  {Doblas}}, \bibinfo {author} {\bibfnamefont {L.}~\bibnamefont
  {Moreno-Ramírez}}, \bibinfo {author} {\bibfnamefont {V.}~\bibnamefont
  {Franco}}, \bibinfo {author} {\bibfnamefont {A.}~\bibnamefont {Conde}},
  \bibinfo {author} {\bibfnamefont {A.}~\bibnamefont {Svalov}},\ and\ \bibinfo
  {author} {\bibfnamefont {G.}~\bibnamefont {Kurlyandskaya}},\ }\bibfield
  {title} {\bibinfo {title} {Nanostructuring as a procedure to control the
  field dependence of the magnetocaloric effect},\ }\href
  {https://doi.org/10.1016/j.matdes.2016.11.085} {\bibfield  {journal}
  {\bibinfo  {journal} {Materials \& Design}\ }\textbf {\bibinfo {volume}
  {114}},\ \bibinfo {pages} {214} (\bibinfo {year} {2017})}\BibitemShut
  {NoStop}%
\bibitem [{\citenamefont {Mukherjee}\ \emph {et~al.}(2009)\citenamefont
  {Mukherjee}, \citenamefont {Sahoo}, \citenamefont {Skomski}, \citenamefont
  {Sellmyer},\ and\ \citenamefont {Binek}}]{Mukherjee2009Magnetocaloric}%
  \BibitemOpen
  \bibfield  {author} {\bibinfo {author} {\bibfnamefont {T.}~\bibnamefont
  {Mukherjee}}, \bibinfo {author} {\bibfnamefont {S.}~\bibnamefont {Sahoo}},
  \bibinfo {author} {\bibfnamefont {R.}~\bibnamefont {Skomski}}, \bibinfo
  {author} {\bibfnamefont {D.~J.}\ \bibnamefont {Sellmyer}},\ and\ \bibinfo
  {author} {\bibfnamefont {C.}~\bibnamefont {Binek}},\ }\bibfield  {title}
  {\bibinfo {title} {Magnetocaloric properties of $\text{Co}/\text{Cr}$
  superlattices},\ }\href {https://doi.org/10.1103/PhysRevB.79.144406}
  {\bibfield  {journal} {\bibinfo  {journal} {Physical Review B}\ }\textbf
  {\bibinfo {volume} {79}},\ \bibinfo {pages} {144406} (\bibinfo {year}
  {2009})}\BibitemShut {NoStop}%
\bibitem [{\citenamefont {Vorobiov}\ \emph {et~al.}(2019)\citenamefont
  {Vorobiov}, \citenamefont {Tomasova}, \citenamefont {Girman}, \citenamefont
  {You}, \citenamefont {Čižmár}, \citenamefont {Orendáč},\ and\
  \citenamefont {Komanicky}}]{Vorobiov2019Optimization}%
  \BibitemOpen
  \bibfield  {author} {\bibinfo {author} {\bibfnamefont {S.}~\bibnamefont
  {Vorobiov}}, \bibinfo {author} {\bibfnamefont {D.}~\bibnamefont {Tomasova}},
  \bibinfo {author} {\bibfnamefont {V.}~\bibnamefont {Girman}}, \bibinfo
  {author} {\bibfnamefont {H.}~\bibnamefont {You}}, \bibinfo {author}
  {\bibfnamefont {E.}~\bibnamefont {Čižmár}}, \bibinfo {author}
  {\bibfnamefont {M.}~\bibnamefont {Orendáč}},\ and\ \bibinfo {author}
  {\bibfnamefont {V.}~\bibnamefont {Komanicky}},\ }\bibfield  {title} {\bibinfo
  {title} {Optimization of the magnetocaloric effect in arrays of
  $\text{Ni}_3\text{Pt}$ nanomagnets},\ }\href
  {https://doi.org/10.1016/j.jmmm.2018.10.137} {\bibfield  {journal} {\bibinfo
  {journal} {Journal of Magnetism and Magnetic Materials}\ }\textbf {\bibinfo
  {volume} {474}},\ \bibinfo {pages} {63} (\bibinfo {year} {2019})}\BibitemShut
  {NoStop}%
\bibitem [{\citenamefont {Barman}\ \emph {et~al.}(2020)\citenamefont {Barman},
  \citenamefont {Chatterjee}, \citenamefont {Dey}, \citenamefont {Datta},\ and\
  \citenamefont {Mukherjee}}]{Barman2020Interface-induced}%
  \BibitemOpen
  \bibfield  {author} {\bibinfo {author} {\bibfnamefont {A.}~\bibnamefont
  {Barman}}, \bibinfo {author} {\bibfnamefont {S.}~\bibnamefont {Chatterjee}},
  \bibinfo {author} {\bibfnamefont {J.~K.}\ \bibnamefont {Dey}}, \bibinfo
  {author} {\bibfnamefont {A.}~\bibnamefont {Datta}},\ and\ \bibinfo {author}
  {\bibfnamefont {D.}~\bibnamefont {Mukherjee}},\ }\bibfield  {title} {\bibinfo
  {title} {Interface-induced enhanced magnetocaloric effect in an epitaxial
  $\text{CoFe}_{2}\text{O}_{4}/\text{La}_{0.7}\text{Sr}_{0.3}\text{MnO}_3$
  heterostructure},\ }\href {https://doi.org/10.1103/PhysRevB.102.054433}
  {\bibfield  {journal} {\bibinfo  {journal} {Physical Review B}\ }\textbf
  {\bibinfo {volume} {102}},\ \bibinfo {pages} {054433} (\bibinfo {year}
  {2020})}\BibitemShut {NoStop}%
\bibitem [{\citenamefont {Khattak}\ \emph {et~al.}(2015)\citenamefont
  {Khattak}, \citenamefont {Aslani}, \citenamefont {Nwokoye}, \citenamefont
  {Siddique}, \citenamefont {Bennett},\ and\ \citenamefont
  {Torre}}]{doi:10.1080/23311916.2015.1050324}%
  \BibitemOpen
  \bibfield  {author} {\bibinfo {author} {\bibfnamefont {K.~S.}\ \bibnamefont
  {Khattak}}, \bibinfo {author} {\bibfnamefont {A.}~\bibnamefont {Aslani}},
  \bibinfo {author} {\bibfnamefont {C.~A.}\ \bibnamefont {Nwokoye}}, \bibinfo
  {author} {\bibfnamefont {A.}~\bibnamefont {Siddique}}, \bibinfo {author}
  {\bibfnamefont {L.~H.}\ \bibnamefont {Bennett}},\ and\ \bibinfo {author}
  {\bibfnamefont {E.~D.}\ \bibnamefont {Torre}},\ }\bibfield  {title} {\bibinfo
  {title} {Magnetocaloric properties of metallic nanostructures},\ }\href
  {https://doi.org/10.1080/23311916.2015.1050324} {\bibfield  {journal}
  {\bibinfo  {journal} {Cogent Engineering}\ }\textbf {\bibinfo {volume} {2}},\
  \bibinfo {pages} {1050324} (\bibinfo {year} {2015})},\ \Eprint
  {https://arxiv.org/abs/https://doi.org/10.1080/23311916.2015.1050324}
  {https://doi.org/10.1080/23311916.2015.1050324} \BibitemShut {NoStop}%
\bibitem [{\citenamefont {Kravets}\ \emph {et~al.}(2012)\citenamefont
  {Kravets}, \citenamefont {Timoshevskii}, \citenamefont {Yanchitsky},
  \citenamefont {Bergmann}, \citenamefont {Buhler}, \citenamefont {Andersson},\
  and\ \citenamefont {Korenivski}}]{Kravets2012Temperature-controlled}%
  \BibitemOpen
  \bibfield  {author} {\bibinfo {author} {\bibfnamefont {A.~F.}\ \bibnamefont
  {Kravets}}, \bibinfo {author} {\bibfnamefont {A.~N.}\ \bibnamefont
  {Timoshevskii}}, \bibinfo {author} {\bibfnamefont {B.~Z.}\ \bibnamefont
  {Yanchitsky}}, \bibinfo {author} {\bibfnamefont {M.~A.}\ \bibnamefont
  {Bergmann}}, \bibinfo {author} {\bibfnamefont {J.}~\bibnamefont {Buhler}},
  \bibinfo {author} {\bibfnamefont {S.}~\bibnamefont {Andersson}},\ and\
  \bibinfo {author} {\bibfnamefont {V.}~\bibnamefont {Korenivski}},\ }\bibfield
   {title} {\bibinfo {title} {Temperature-controlled interlayer exchange
  coupling in strong/weak ferromagnetic multilayers: A thermomagnetic curie
  switch},\ }\href {https://doi.org/10.1103/PhysRevB.86.214413} {\bibfield
  {journal} {\bibinfo  {journal} {Physical Review B}\ }\textbf {\bibinfo
  {volume} {86}},\ \bibinfo {pages} {214413} (\bibinfo {year}
  {2012})}\BibitemShut {NoStop}%
\bibitem [{\citenamefont {Kravets}\ \emph {et~al.}(2014)\citenamefont
  {Kravets}, \citenamefont {Dzhezherya}, \citenamefont {Tovstolytkin},
  \citenamefont {Kozak}, \citenamefont {Gryshchuk}, \citenamefont {Savina},
  \citenamefont {Pashchenko}, \citenamefont {Gnatchenko}, \citenamefont
  {Koop},\ and\ \citenamefont {Korenivski}}]{Kravets2014Synthetic}%
  \BibitemOpen
  \bibfield  {author} {\bibinfo {author} {\bibfnamefont {A.~F.}\ \bibnamefont
  {Kravets}}, \bibinfo {author} {\bibfnamefont {Y.~I.}\ \bibnamefont
  {Dzhezherya}}, \bibinfo {author} {\bibfnamefont {A.~I.}\ \bibnamefont
  {Tovstolytkin}}, \bibinfo {author} {\bibfnamefont {I.~M.}\ \bibnamefont
  {Kozak}}, \bibinfo {author} {\bibfnamefont {A.}~\bibnamefont {Gryshchuk}},
  \bibinfo {author} {\bibfnamefont {Y.~O.}\ \bibnamefont {Savina}}, \bibinfo
  {author} {\bibfnamefont {V.~A.}\ \bibnamefont {Pashchenko}}, \bibinfo
  {author} {\bibfnamefont {S.~L.}\ \bibnamefont {Gnatchenko}}, \bibinfo
  {author} {\bibfnamefont {B.}~\bibnamefont {Koop}},\ and\ \bibinfo {author}
  {\bibfnamefont {V.}~\bibnamefont {Korenivski}},\ }\bibfield  {title}
  {\bibinfo {title} {Synthetic ferrimagnets with thermomagnetic switching},\
  }\href {https://doi.org/10.1103/PhysRevB.90.104427} {\bibfield  {journal}
  {\bibinfo  {journal} {Physical Review B}\ }\textbf {\bibinfo {volume} {90}},\
  \bibinfo {pages} {104427} (\bibinfo {year} {2014})}\BibitemShut {NoStop}%
\bibitem [{\citenamefont {Kravets}\ \emph {et~al.}(2016)\citenamefont
  {Kravets}, \citenamefont {Polishchuk}, \citenamefont {Dzhezherya},
  \citenamefont {Tovstolytkin}, \citenamefont {Golub},\ and\ \citenamefont
  {Korenivski}}]{Kravets_2016}%
  \BibitemOpen
  \bibfield  {author} {\bibinfo {author} {\bibfnamefont {A.~F.}\ \bibnamefont
  {Kravets}}, \bibinfo {author} {\bibfnamefont {D.~M.}\ \bibnamefont
  {Polishchuk}}, \bibinfo {author} {\bibfnamefont {Y.~I.}\ \bibnamefont
  {Dzhezherya}}, \bibinfo {author} {\bibfnamefont {A.~I.}\ \bibnamefont
  {Tovstolytkin}}, \bibinfo {author} {\bibfnamefont {V.~O.}\ \bibnamefont
  {Golub}},\ and\ \bibinfo {author} {\bibfnamefont {V.}~\bibnamefont
  {Korenivski}},\ }\bibfield  {title} {\bibinfo {title} {Anisotropic
  magnetization relaxation in ferromagnetic multilayers with variable
  interlayer exchange coupling},\ }\bibfield  {journal} {\bibinfo  {journal}
  {Physical Review B}\ }\textbf {\bibinfo {volume} {94}},\ \href
  {https://doi.org/https://doi.org/10.1103/PhysRevB.94.064429}
  {https://doi.org/10.1103/PhysRevB.94.064429} (\bibinfo {year}
  {2016})\BibitemShut {NoStop}%
\bibitem [{\citenamefont {Fraerman}\ and\ \citenamefont
  {Shereshevskii}(2015)}]{Fraerman2015Magnetocaloric}%
  \BibitemOpen
  \bibfield  {author} {\bibinfo {author} {\bibfnamefont {A.~A.}\ \bibnamefont
  {Fraerman}}\ and\ \bibinfo {author} {\bibfnamefont {I.~A.}\ \bibnamefont
  {Shereshevskii}},\ }\bibfield  {title} {\bibinfo {title} {Magnetocaloric
  effect in ferromagnet/paramagnet multilayer structures},\ }\href
  {https://doi.org/10.1134/S0021364015090088} {\bibfield  {journal} {\bibinfo
  {journal} {JETP Letters}\ }\textbf {\bibinfo {volume} {101}},\ \bibinfo
  {pages} {618} (\bibinfo {year} {2015})}\BibitemShut {NoStop}%
\bibitem [{\citenamefont {Vdovichev}\ \emph {et~al.}(2018)\citenamefont
  {Vdovichev}, \citenamefont {Polushkin}, \citenamefont {Rodionov},
  \citenamefont {Prudnikov}, \citenamefont {Chang},\ and\ \citenamefont
  {Fraerman}}]{Vdovichev2018High}%
  \BibitemOpen
  \bibfield  {author} {\bibinfo {author} {\bibfnamefont {S.~N.}\ \bibnamefont
  {Vdovichev}}, \bibinfo {author} {\bibfnamefont {N.~I.}\ \bibnamefont
  {Polushkin}}, \bibinfo {author} {\bibfnamefont {I.~D.}\ \bibnamefont
  {Rodionov}}, \bibinfo {author} {\bibfnamefont {V.~N.}\ \bibnamefont
  {Prudnikov}}, \bibinfo {author} {\bibfnamefont {J.}~\bibnamefont {Chang}},\
  and\ \bibinfo {author} {\bibfnamefont {A.~A.}\ \bibnamefont {Fraerman}},\
  }\bibfield  {title} {\bibinfo {title} {High magnetocaloric efficiency of a
  $\text{NiFe/NiCu/CoFe/MnIr}$ multilayer in a small magnetic field},\ }\href
  {https://doi.org/10.1103/PhysRevB.98.014428} {\bibfield  {journal} {\bibinfo
  {journal} {Physical Review B}\ }\textbf {\bibinfo {volume} {98}},\ \bibinfo
  {pages} {014428} (\bibinfo {year} {2018})}\BibitemShut {NoStop}%
\bibitem [{\citenamefont {Polushkin}\ \emph {et~al.}(2019)\citenamefont
  {Polushkin}, \citenamefont {Pashenkin}, \citenamefont {Fadeev}, \citenamefont
  {Lähderanta},\ and\ \citenamefont {Fraerman}}]{Polushkin2019Magnetic}%
  \BibitemOpen
  \bibfield  {author} {\bibinfo {author} {\bibfnamefont {N.}~\bibnamefont
  {Polushkin}}, \bibinfo {author} {\bibfnamefont {I.}~\bibnamefont
  {Pashenkin}}, \bibinfo {author} {\bibfnamefont {E.}~\bibnamefont {Fadeev}},
  \bibinfo {author} {\bibfnamefont {E.}~\bibnamefont {Lähderanta}},\ and\
  \bibinfo {author} {\bibfnamefont {A.}~\bibnamefont {Fraerman}},\ }\bibfield
  {title} {\bibinfo {title} {Magnetic and magnetocaloric properties of
  $\text{Py/Gd/CoFe/IrMn}$ stacks},\ }\href
  {https://doi.org/10.1016/j.jmmm.2019.165601} {\bibfield  {journal} {\bibinfo
  {journal} {Journal of Magnetism and Magnetic Materials}\ }\textbf {\bibinfo
  {volume} {491}},\ \bibinfo {pages} {165601} (\bibinfo {year}
  {2019})}\BibitemShut {NoStop}%
\bibitem [{\citenamefont {Kuznetsov}\ \emph {et~al.}(2020)\citenamefont
  {Kuznetsov}, \citenamefont {Pashenkin}, \citenamefont {Polushkin},
  \citenamefont {Sapozhnikov},\ and\ \citenamefont
  {Fraerman}}]{Kuznetsov2020Magnetocaloric}%
  \BibitemOpen
  \bibfield  {author} {\bibinfo {author} {\bibfnamefont {M.~A.}\ \bibnamefont
  {Kuznetsov}}, \bibinfo {author} {\bibfnamefont {I.~Y.}\ \bibnamefont
  {Pashenkin}}, \bibinfo {author} {\bibfnamefont {N.~I.}\ \bibnamefont
  {Polushkin}}, \bibinfo {author} {\bibfnamefont {M.~V.}\ \bibnamefont
  {Sapozhnikov}},\ and\ \bibinfo {author} {\bibfnamefont {A.~A.}\ \bibnamefont
  {Fraerman}},\ }\bibfield  {title} {\bibinfo {title} {Magnetocaloric effect in
  exchange-coupled strong/weak/strong ferromagnet stacks},\ }\href
  {https://doi.org/10.1063/5.0003223} {\bibfield  {journal} {\bibinfo
  {journal} {Journal of Applied Physics}\ }\textbf {\bibinfo {volume} {127}},\
  \bibinfo {pages} {183904} (\bibinfo {year} {2020})}\BibitemShut {NoStop}%
\bibitem [{\citenamefont {Kuznetsov}\ \emph {et~al.}(2021)\citenamefont
  {Kuznetsov}, \citenamefont {Drovosekov},\ and\ \citenamefont
  {Fraerman}}]{Kuznetsov2021Magnetocaloric}%
  \BibitemOpen
  \bibfield  {author} {\bibinfo {author} {\bibfnamefont {M.~A.}\ \bibnamefont
  {Kuznetsov}}, \bibinfo {author} {\bibfnamefont {A.~B.}\ \bibnamefont
  {Drovosekov}},\ and\ \bibinfo {author} {\bibfnamefont {A.~A.}\ \bibnamefont
  {Fraerman}},\ }\bibfield  {title} {\bibinfo {title} {Magnetocaloric effect in
  nanosystems based on ferromagnets with different curie temperatures},\ }\href
  {https://doi.org/10.1134/S1063776121010131} {\bibfield  {journal} {\bibinfo
  {journal} {Journal of Experimental and Theoretical Physics}\ }\textbf
  {\bibinfo {volume} {132}},\ \bibinfo {pages} {79} (\bibinfo {year}
  {2021})}\BibitemShut {NoStop}%
\bibitem [{\citenamefont {Polishchuk}\ \emph {et~al.}(2018)\citenamefont
  {Polishchuk}, \citenamefont {Tykhonenko-Polishchuk}, \citenamefont
  {Holmgren}, \citenamefont {Kravets}, \citenamefont {Tovstolytkin},\ and\
  \citenamefont {Korenivski}}]{Polishchuk2018Giant}%
  \BibitemOpen
  \bibfield  {author} {\bibinfo {author} {\bibfnamefont {D.~M.}\ \bibnamefont
  {Polishchuk}}, \bibinfo {author} {\bibfnamefont {Y.~O.}\ \bibnamefont
  {Tykhonenko-Polishchuk}}, \bibinfo {author} {\bibfnamefont {E.}~\bibnamefont
  {Holmgren}}, \bibinfo {author} {\bibfnamefont {A.~F.}\ \bibnamefont
  {Kravets}}, \bibinfo {author} {\bibfnamefont {A.~I.}\ \bibnamefont
  {Tovstolytkin}},\ and\ \bibinfo {author} {\bibfnamefont {V.}~\bibnamefont
  {Korenivski}},\ }\bibfield  {title} {\bibinfo {title} {Giant magnetocaloric
  effect driven by indirect exchange in magnetic multilayers},\ }\href
  {https://doi.org/10.1103/PhysRevMaterials.2.114402} {\bibfield  {journal}
  {\bibinfo  {journal} {Physical Review Materials}\ }\textbf {\bibinfo {volume}
  {2}},\ \bibinfo {pages} {114402} (\bibinfo {year} {2018})}\BibitemShut
  {NoStop}%
\bibitem [{\citenamefont {Lim}\ \emph {et~al.}(2013)\citenamefont {Lim},
  \citenamefont {Ebrahim-Zadeh}, \citenamefont {Owens}, \citenamefont
  {Hentschel},\ and\ \citenamefont {Urazhdin}}]{Lim2013Temperature-dependent}%
  \BibitemOpen
  \bibfield  {author} {\bibinfo {author} {\bibfnamefont {W.~L.}\ \bibnamefont
  {Lim}}, \bibinfo {author} {\bibfnamefont {N.}~\bibnamefont {Ebrahim-Zadeh}},
  \bibinfo {author} {\bibfnamefont {J.~C.}\ \bibnamefont {Owens}}, \bibinfo
  {author} {\bibfnamefont {H.~G.~E.}\ \bibnamefont {Hentschel}},\ and\ \bibinfo
  {author} {\bibfnamefont {S.}~\bibnamefont {Urazhdin}},\ }\bibfield  {title}
  {\bibinfo {title} {Temperature-dependent proximity magnetism in
  $\text{Pt}$},\ }\href {https://doi.org/10.1063/1.4802954} {\bibfield
  {journal} {\bibinfo  {journal} {Applied Physics Letters}\ }\textbf {\bibinfo
  {volume} {102}},\ \bibinfo {pages} {162404} (\bibinfo {year}
  {2013})}\BibitemShut {NoStop}%
\bibitem [{\citenamefont {Song}\ \emph {et~al.}(2011)\citenamefont {Song},
  \citenamefont {Sperl}, \citenamefont {Utz}, \citenamefont {Ciorga},
  \citenamefont {Woltersdorf}, \citenamefont {Schuh}, \citenamefont {Bougeard},
  \citenamefont {Back},\ and\ \citenamefont {Weiss}}]{Song2011Proximity}%
  \BibitemOpen
  \bibfield  {author} {\bibinfo {author} {\bibfnamefont {C.}~\bibnamefont
  {Song}}, \bibinfo {author} {\bibfnamefont {M.}~\bibnamefont {Sperl}},
  \bibinfo {author} {\bibfnamefont {M.}~\bibnamefont {Utz}}, \bibinfo {author}
  {\bibfnamefont {M.}~\bibnamefont {Ciorga}}, \bibinfo {author} {\bibfnamefont
  {G.}~\bibnamefont {Woltersdorf}}, \bibinfo {author} {\bibfnamefont
  {D.}~\bibnamefont {Schuh}}, \bibinfo {author} {\bibfnamefont
  {D.}~\bibnamefont {Bougeard}}, \bibinfo {author} {\bibfnamefont {C.~H.}\
  \bibnamefont {Back}},\ and\ \bibinfo {author} {\bibfnamefont
  {D.}~\bibnamefont {Weiss}},\ }\bibfield  {title} {\bibinfo {title} {Proximity
  induced enhancement of the curie temperature in hybrid spin injection
  devices},\ }\href {https://doi.org/10.1103/PhysRevLett.107.056601} {\bibfield
   {journal} {\bibinfo  {journal} {Physical Review Letters}\ }\textbf {\bibinfo
  {volume} {107}},\ \bibinfo {pages} {056601} (\bibinfo {year}
  {2011})}\BibitemShut {NoStop}%
\bibitem [{\citenamefont {Camley}(1987)}]{Camley1987Surface}%
  \BibitemOpen
  \bibfield  {author} {\bibinfo {author} {\bibfnamefont {R.~E.}\ \bibnamefont
  {Camley}},\ }\bibfield  {title} {\bibinfo {title} {Surface spin reorientation
  in thin gd films on fe in an applied magnetic field},\ }\href
  {https://doi.org/10.1103/PhysRevB.35.3608} {\bibfield  {journal} {\bibinfo
  {journal} {Physical Review B}\ }\textbf {\bibinfo {volume} {35}},\ \bibinfo
  {pages} {3608} (\bibinfo {year} {1987})}\BibitemShut {NoStop}%
\bibitem [{\citenamefont {Camley}\ and\ \citenamefont
  {Tilley}(1988)}]{Camley1988Phase}%
  \BibitemOpen
  \bibfield  {author} {\bibinfo {author} {\bibfnamefont {R.~E.}\ \bibnamefont
  {Camley}}\ and\ \bibinfo {author} {\bibfnamefont {D.~R.}\ \bibnamefont
  {Tilley}},\ }\bibfield  {title} {\bibinfo {title} {Phase transitions in
  magnetic superlattices},\ }\href {https://doi.org/10.1103/PhysRevB.37.3413}
  {\bibfield  {journal} {\bibinfo  {journal} {Physical Review B}\ }\textbf
  {\bibinfo {volume} {37}},\ \bibinfo {pages} {3413} (\bibinfo {year}
  {1988})}\BibitemShut {NoStop}%
\bibitem [{\citenamefont {Camley}(1989)}]{Camley1989Properties}%
  \BibitemOpen
  \bibfield  {author} {\bibinfo {author} {\bibfnamefont {R.~E.}\ \bibnamefont
  {Camley}},\ }\bibfield  {title} {\bibinfo {title} {Properties of magnetic
  superlattices with antiferromagnetic interfacial coupling: Magnetization,
  susceptibility, and compensation points},\ }\href
  {https://doi.org/10.1103/PhysRevB.39.12316} {\bibfield  {journal} {\bibinfo
  {journal} {Physical Review B}\ }\textbf {\bibinfo {volume} {39}},\ \bibinfo
  {pages} {12316} (\bibinfo {year} {1989})}\BibitemShut {NoStop}%
\bibitem [{\citenamefont {Magnus}\ \emph {et~al.}(2016)\citenamefont {Magnus},
  \citenamefont {Brooks-Bartlett}, \citenamefont {Moubah}, \citenamefont
  {Procter}, \citenamefont {Andersson}, \citenamefont {Hase}, \citenamefont
  {Banks},\ and\ \citenamefont {Hjörvarsson}}]{Magnus2016Long-range}%
  \BibitemOpen
  \bibfield  {author} {\bibinfo {author} {\bibfnamefont {F.}~\bibnamefont
  {Magnus}}, \bibinfo {author} {\bibfnamefont {M.~E.}\ \bibnamefont
  {Brooks-Bartlett}}, \bibinfo {author} {\bibfnamefont {R.}~\bibnamefont
  {Moubah}}, \bibinfo {author} {\bibfnamefont {R.~A.}\ \bibnamefont {Procter}},
  \bibinfo {author} {\bibfnamefont {G.}~\bibnamefont {Andersson}}, \bibinfo
  {author} {\bibfnamefont {T.~P.~A.}\ \bibnamefont {Hase}}, \bibinfo {author}
  {\bibfnamefont {S.~T.}\ \bibnamefont {Banks}},\ and\ \bibinfo {author}
  {\bibfnamefont {B.}~\bibnamefont {Hjörvarsson}},\ }\bibfield  {title}
  {\bibinfo {title} {Long-range magnetic interactions and proximity effects in
  an amorphous exchange-spring magnet},\ }\href
  {https://doi.org/10.1038/ncomms11931} {\bibfield  {journal} {\bibinfo
  {journal} {Nature Communications}\ }\textbf {\bibinfo {volume} {7}},\
  \bibinfo {pages} {ncomms11931} (\bibinfo {year} {2016})}\BibitemShut
  {NoStop}%
\bibitem [{\citenamefont {Ravi~Kumar}\ and\ \citenamefont
  {Kaul}(2018)}]{RaviKumar2018Thickness}%
  \BibitemOpen
  \bibfield  {author} {\bibinfo {author} {\bibfnamefont {B.}~\bibnamefont
  {Ravi~Kumar}}\ and\ \bibinfo {author} {\bibfnamefont {S.}~\bibnamefont
  {Kaul}},\ }\bibfield  {title} {\bibinfo {title} {Thickness as a control
  parameter for magnetocaloric effect in $\text{Cr}_{75-x}\text{Fe}_{25+x}$ (x
  = 0, 5) thin films},\ }\href {https://doi.org/10.1016/j.jmmm.2018.04.011}
  {\bibfield  {journal} {\bibinfo  {journal} {Journal of Magnetism and Magnetic
  Materials}\ }\textbf {\bibinfo {volume} {460}},\ \bibinfo {pages} {312}
  (\bibinfo {year} {2018})}\BibitemShut {NoStop}%
\bibitem [{\citenamefont {Ravi~Kumar}\ and\ \citenamefont
  {Kaul}(2015)}]{RaviKumar2015Magnetic}%
  \BibitemOpen
  \bibfield  {author} {\bibinfo {author} {\bibfnamefont {B.}~\bibnamefont
  {Ravi~Kumar}}\ and\ \bibinfo {author} {\bibfnamefont {S.}~\bibnamefont
  {Kaul}},\ }\bibfield  {title} {\bibinfo {title} {Magnetic order-disorder
  phase transition in $\text{Cr}_{70}\text{Fe}_{30}$ thin films},\ }\href
  {https://doi.org/10.1016/j.jallcom.2015.08.173} {\bibfield  {journal}
  {\bibinfo  {journal} {Journal of Alloys and Compounds}\ }\textbf {\bibinfo
  {volume} {652}},\ \bibinfo {pages} {479} (\bibinfo {year}
  {2015})}\BibitemShut {NoStop}%
\bibitem [{\citenamefont {Tagliati}\ \emph {et~al.}(2009)\citenamefont
  {Tagliati}, \citenamefont {Rydh}, \citenamefont {Xie}, \citenamefont {Welp},\
  and\ \citenamefont {Kwok}}]{Tagliati2009Membrane-based}%
  \BibitemOpen
  \bibfield  {author} {\bibinfo {author} {\bibfnamefont {S.}~\bibnamefont
  {Tagliati}}, \bibinfo {author} {\bibfnamefont {A.}~\bibnamefont {Rydh}},
  \bibinfo {author} {\bibfnamefont {R.}~\bibnamefont {Xie}}, \bibinfo {author}
  {\bibfnamefont {U.}~\bibnamefont {Welp}},\ and\ \bibinfo {author}
  {\bibfnamefont {W.~K.}\ \bibnamefont {Kwok}},\ }\bibfield  {title} {\bibinfo
  {title} {Membrane-based calorimetry for studies of sub-microgram samples},\
  }\href {https://doi.org/10.1088/1742-6596/150/5/052256} {\bibfield  {journal}
  {\bibinfo  {journal} {Journal of Physics: Conference Series}\ }\textbf
  {\bibinfo {volume} {150}},\ \bibinfo {pages} {052256} (\bibinfo {year}
  {2009})}\BibitemShut {NoStop}%
\bibitem [{\citenamefont {van Herwaarden}(2005)}]{Herwaarden2005Overview}%
  \BibitemOpen
  \bibfield  {author} {\bibinfo {author} {\bibfnamefont {A.}~\bibnamefont {van
  Herwaarden}},\ }\bibfield  {title} {\bibinfo {title} {Overview of calorimeter
  chips for various applications},\ }\href
  {https://doi.org/10.1016/j.tca.2005.04.027} {\bibfield  {journal} {\bibinfo
  {journal} {Thermochimica Acta}\ }\textbf {\bibinfo {volume} {432}},\ \bibinfo
  {pages} {192} (\bibinfo {year} {2005})}\BibitemShut {NoStop}%
\bibitem [{\citenamefont {Goodman}\ \emph {et~al.}(1998)\citenamefont
  {Goodman}, \citenamefont {O’Grady},\ and\ \citenamefont
  {Walmsley}}]{GoodmanIEEE}%
  \BibitemOpen
  \bibfield  {author} {\bibinfo {author} {\bibfnamefont {M.}~\bibnamefont
  {Goodman}}, \bibinfo {author} {\bibfnamefont {K.}~\bibnamefont {O’Grady}},\
  and\ \bibinfo {author} {\bibfnamefont {N.~S.}\ \bibnamefont {Walmsley}},\
  }\bibfield  {title} {\bibinfo {title} {Magnetisation reversal in spin-valve
  structures},\ }\href {https://doi.org/10.1109/20.617792} {\bibfield
  {journal} {\bibinfo  {journal} {IEEE TRANSACTIONS ON MAGENTICS}\ }\textbf
  {\bibinfo {volume} {33}},\ \bibinfo {pages} {3} (\bibinfo {year}
  {1998})}\BibitemShut {NoStop}%
\bibitem [{\citenamefont {Ledue}\ \emph {et~al.}(2014)\citenamefont {Ledue},
  \citenamefont {Maitre}, \citenamefont {Barbe},\ and\ \citenamefont
  {Lechevallier}}]{Ledue2014Temperature}%
  \BibitemOpen
  \bibfield  {author} {\bibinfo {author} {\bibfnamefont {D.}~\bibnamefont
  {Ledue}}, \bibinfo {author} {\bibfnamefont {A.}~\bibnamefont {Maitre}},
  \bibinfo {author} {\bibfnamefont {F.}~\bibnamefont {Barbe}},\ and\ \bibinfo
  {author} {\bibfnamefont {L.}~\bibnamefont {Lechevallier}},\ }\bibfield
  {title} {\bibinfo {title} {Temperature dependence of the exchange bias
  properties of ferromagnetic/antiferromagnetic polycrystalline bilayers},\
  }\href {https://doi.org/10.1016/j.jmmm.2014.07.021} {\bibfield  {journal}
  {\bibinfo  {journal} {Journal of Magnetism and Magnetic Materials}\ }\textbf
  {\bibinfo {volume} {372}},\ \bibinfo {pages} {134} (\bibinfo {year}
  {2014})}\BibitemShut {NoStop}%
\bibitem [{\citenamefont {Nogu\'{e}s}\ and\ \citenamefont
  {Schuller}(1999)}]{Nogues1999Exchange}%
  \BibitemOpen
  \bibfield  {author} {\bibinfo {author} {\bibfnamefont {J.}~\bibnamefont
  {Nogu\'{e}s}}\ and\ \bibinfo {author} {\bibfnamefont {I.~K.}\ \bibnamefont
  {Schuller}},\ }\bibfield  {title} {\bibinfo {title} {Exchange bias},\ }\href
  {https://doi.org/10.1016/S0304-8853(98)00266-2} {\bibfield  {journal}
  {\bibinfo  {journal} {Journal of Magnetism and Magnetic Materials}\ }\textbf
  {\bibinfo {volume} {192}},\ \bibinfo {pages} {203} (\bibinfo {year}
  {1999})}\BibitemShut {NoStop}%
\bibitem [{\citenamefont {Jenkins}\ \emph {et~al.}(2020)\citenamefont
  {Jenkins}, \citenamefont {Fan}, \citenamefont {Gaina}, \citenamefont
  {Chantrell}, \citenamefont {Klemmer},\ and\ \citenamefont
  {Evans}}]{PhysRevB.102.140404}%
  \BibitemOpen
  \bibfield  {author} {\bibinfo {author} {\bibfnamefont {S.}~\bibnamefont
  {Jenkins}}, \bibinfo {author} {\bibfnamefont {W.~J.}\ \bibnamefont {Fan}},
  \bibinfo {author} {\bibfnamefont {R.}~\bibnamefont {Gaina}}, \bibinfo
  {author} {\bibfnamefont {R.~W.}\ \bibnamefont {Chantrell}}, \bibinfo {author}
  {\bibfnamefont {T.}~\bibnamefont {Klemmer}},\ and\ \bibinfo {author}
  {\bibfnamefont {R.~F.~L.}\ \bibnamefont {Evans}},\ }\bibfield  {title}
  {\bibinfo {title} {Atomistic origin of exchange anisotropy in noncollinear
  $\gamma-\text{IrMn}_{3}-\text{CoFe}$ bilayers},\ }\href
  {https://doi.org/10.1103/PhysRevB.102.140404} {\bibfield  {journal} {\bibinfo
   {journal} {Phys. Rev. B}\ }\textbf {\bibinfo {volume} {102}},\ \bibinfo
  {pages} {140404} (\bibinfo {year} {2020})}\BibitemShut {NoStop}%
\bibitem [{Note1()}]{Note1}%
  \BibitemOpen
  \bibinfo {note} {We note that using the indirect Maxwell-relation approach
  with the magnetization data for the trilayer sample yielded MCE, which did
  not correlate with the direct measurements either in magnitude or temperature
  dependence, and was discarded from further discussion and
  analysis.}\BibitemShut {Stop}%
\bibitem [{\citenamefont {Berger}(1996)}]{PhysRevB.54.9353}%
  \BibitemOpen
  \bibfield  {author} {\bibinfo {author} {\bibfnamefont {L.}~\bibnamefont
  {Berger}},\ }\bibfield  {title} {\bibinfo {title} {Emission of spin waves by
  a magnetic multilayer traversed by a current},\ }\href
  {https://doi.org/10.1103/PhysRevB.54.9353} {\bibfield  {journal} {\bibinfo
  {journal} {Phys. Rev. B}\ }\textbf {\bibinfo {volume} {54}},\ \bibinfo
  {pages} {9353} (\bibinfo {year} {1996})}\BibitemShut {NoStop}%
\bibitem [{\citenamefont {Gorobets}\ \emph {et~al.}(2000)\citenamefont
  {Gorobets}, \citenamefont {Dzhezherya},\ and\ \citenamefont
  {Kravets}}]{Gorobets2000}%
  \BibitemOpen
  \bibfield  {author} {\bibinfo {author} {\bibfnamefont {Y.~I.}\ \bibnamefont
  {Gorobets}}, \bibinfo {author} {\bibfnamefont {Y.~I.}\ \bibnamefont
  {Dzhezherya}},\ and\ \bibinfo {author} {\bibfnamefont {A.~F.}\ \bibnamefont
  {Kravets}},\ }\bibfield  {title} {\bibinfo {title} {Magnetic ordering in
  granular system},\ }\href {https://doi.org/https://doi.org/10.1134/1.1131179}
  {\bibfield  {journal} {\bibinfo  {journal} {Physics of the Solid State}\
  }\textbf {\bibinfo {volume} {42}},\ \bibinfo {pages} {121} (\bibinfo {year}
  {2000})}\BibitemShut {NoStop}%
\bibitem [{\citenamefont {Bass}\ and\ \citenamefont
  {Pratt}(2007)}]{Bass2007Spin-diffusion}%
  \BibitemOpen
  \bibfield  {author} {\bibinfo {author} {\bibfnamefont {J.}~\bibnamefont
  {Bass}}\ and\ \bibinfo {author} {\bibfnamefont {W.~P.}\ \bibnamefont
  {Pratt}},\ }\bibfield  {title} {\bibinfo {title} {Spin-diffusion lengths in
  metals and alloys, and spin-flipping at metal/metal interfaces: an
  experimentalist’s critical review},\ }\href
  {https://doi.org/10.1088/0953-8984/19/18/183201} {\bibfield  {journal}
  {\bibinfo  {journal} {Journal of Physics: Condensed Matter}\ }\textbf
  {\bibinfo {volume} {19}},\ \bibinfo {pages} {183201} (\bibinfo {year}
  {2007})}\BibitemShut {NoStop}%
\bibitem [{\citenamefont {Wang}\ \emph {et~al.}(1990)\citenamefont {Wang},
  \citenamefont {Levy},\ and\ \citenamefont {Fry}}]{PhysRevLett.65.2732}%
  \BibitemOpen
  \bibfield  {author} {\bibinfo {author} {\bibfnamefont {Y.}~\bibnamefont
  {Wang}}, \bibinfo {author} {\bibfnamefont {P.~M.}\ \bibnamefont {Levy}},\
  and\ \bibinfo {author} {\bibfnamefont {J.~L.}\ \bibnamefont {Fry}},\
  }\bibfield  {title} {\bibinfo {title} {Interlayer magnetic coupling in
  $\text{Fe/Cr}$ multilayered structures},\ }\href
  {https://doi.org/10.1103/PhysRevLett.65.2732} {\bibfield  {journal} {\bibinfo
   {journal} {Phys. Rev. Lett.}\ }\textbf {\bibinfo {volume} {65}},\ \bibinfo
  {pages} {2732} (\bibinfo {year} {1990})}\BibitemShut {NoStop}%
\bibitem [{\citenamefont {Barthélémy}\ \emph {et~al.}(1990)\citenamefont
  {Barthélémy}, \citenamefont {Fert}, \citenamefont {Baibich}, \citenamefont
  {Hadjoudj}, \citenamefont {Petroff}, \citenamefont {Etienne}, \citenamefont
  {Cabanel}, \citenamefont {Lequien}, \citenamefont {Nguyen Van~Dau},\ and\
  \citenamefont {Creuzet}}]{fert2}%
  \BibitemOpen
  \bibfield  {author} {\bibinfo {author} {\bibfnamefont {A.}~\bibnamefont
  {Barthélémy}}, \bibinfo {author} {\bibfnamefont {A.}~\bibnamefont {Fert}},
  \bibinfo {author} {\bibfnamefont {M.~N.}\ \bibnamefont {Baibich}}, \bibinfo
  {author} {\bibfnamefont {S.}~\bibnamefont {Hadjoudj}}, \bibinfo {author}
  {\bibfnamefont {F.}~\bibnamefont {Petroff}}, \bibinfo {author} {\bibfnamefont
  {P.}~\bibnamefont {Etienne}}, \bibinfo {author} {\bibfnamefont
  {R.}~\bibnamefont {Cabanel}}, \bibinfo {author} {\bibfnamefont
  {S.}~\bibnamefont {Lequien}}, \bibinfo {author} {\bibfnamefont
  {F.}~\bibnamefont {Nguyen Van~Dau}},\ and\ \bibinfo {author} {\bibfnamefont
  {G.}~\bibnamefont {Creuzet}},\ }\bibfield  {title} {\bibinfo {title}
  {Magnetic and transport properties of $\text{Fe/Cr}$ superlattices
  (invited)},\ }\href {https://doi.org/10.1063/1.346013} {\bibfield  {journal}
  {\bibinfo  {journal} {Journal of Applied Physics}\ }\textbf {\bibinfo
  {volume} {67}},\ \bibinfo {pages} {5908} (\bibinfo {year} {1990})},\ \Eprint
  {https://arxiv.org/abs/https://doi.org/10.1063/1.346013}
  {https://doi.org/10.1063/1.346013} \BibitemShut {NoStop}%
\bibitem [{\citenamefont {Parkin}\ \emph {et~al.}(1990)\citenamefont {Parkin},
  \citenamefont {More},\ and\ \citenamefont {Roche}}]{PhysRevLett.64.2304}%
  \BibitemOpen
  \bibfield  {author} {\bibinfo {author} {\bibfnamefont {S.~S.~P.}\
  \bibnamefont {Parkin}}, \bibinfo {author} {\bibfnamefont {N.}~\bibnamefont
  {More}},\ and\ \bibinfo {author} {\bibfnamefont {K.~P.}\ \bibnamefont
  {Roche}},\ }\bibfield  {title} {\bibinfo {title} {Oscillations in exchange
  coupling and magnetoresistance in metallic superlattice structures:
  $\text{Co/Ru}$, $\text{Co/Cr}$, and $\text{Fe/Cr}$},\ }\href
  {https://doi.org/10.1103/PhysRevLett.64.2304} {\bibfield  {journal} {\bibinfo
   {journal} {Phys. Rev. Lett.}\ }\textbf {\bibinfo {volume} {64}},\ \bibinfo
  {pages} {2304} (\bibinfo {year} {1990})}\BibitemShut {NoStop}%
\bibitem [{\citenamefont {Persson}\ \emph {et~al.}(2022)\citenamefont
  {Persson}, \citenamefont {Kulyk}, \citenamefont {Kravets},\ and\
  \citenamefont {Korenivski}}]{Milton_2022}%
  \BibitemOpen
  \bibfield  {author} {\bibinfo {author} {\bibfnamefont {M.}~\bibnamefont
  {Persson}}, \bibinfo {author} {\bibfnamefont {M.}~\bibnamefont {Kulyk}},
  \bibinfo {author} {\bibfnamefont {A.}~\bibnamefont {Kravets}},\ and\ \bibinfo
  {author} {\bibfnamefont {V.}~\bibnamefont {Korenivski}},\ }\bibfield  {title}
  {\bibinfo {title} {Proximity-enhanced magnetocaloric effect in ferromagnetic
  trilayers},\ }\href {https://doi.org/10.48550/arXiv.2208.01379} {\bibfield
  {journal} {\bibinfo  {journal} {arXiv:2208.01379}\ } (\bibinfo {year}
  {2022})}\BibitemShut {NoStop}%
\bibitem [{\citenamefont {Crangle}\ and\ \citenamefont
  {Goodman}(1971)}]{crangle1971magnetization}%
  \BibitemOpen
  \bibfield  {author} {\bibinfo {author} {\bibfnamefont {J.}~\bibnamefont
  {Crangle}}\ and\ \bibinfo {author} {\bibfnamefont {G.}~\bibnamefont
  {Goodman}},\ }\bibfield  {title} {\bibinfo {title} {The magnetization of pure
  iron and nickel},\ }\href
  {https://doi.org/https://doi.org/10.1098/rspa.1971.0044} {\bibfield
  {journal} {\bibinfo  {journal} {Proceedings of the Royal Society of London.
  A. Mathematical and Physical Sciences}\ }\textbf {\bibinfo {volume} {321}},\
  \bibinfo {pages} {477} (\bibinfo {year} {1971})}\BibitemShut {NoStop}%
\bibitem [{\citenamefont {{El Hafidi}}\ \emph {et~al.}(2018)\citenamefont {{El
  Hafidi}}, \citenamefont {Boubekri},\ and\ \citenamefont {{El
  Hafidi}}}]{ELHAFIDI2018500}%
  \BibitemOpen
  \bibfield  {author} {\bibinfo {author} {\bibfnamefont {M.~Y.}\ \bibnamefont
  {{El Hafidi}}}, \bibinfo {author} {\bibfnamefont {A.}~\bibnamefont
  {Boubekri}},\ and\ \bibinfo {author} {\bibfnamefont {M.}~\bibnamefont {{El
  Hafidi}}},\ }\bibfield  {title} {\bibinfo {title} {Calculation model on
  magnetocaloric effect (mce) and relative cooling power (rcp) in composite
  materials at room temperature},\ }\href
  {https://doi.org/https://doi.org/10.1016/j.jmmm.2017.10.083} {\bibfield
  {journal} {\bibinfo  {journal} {Journal of Magnetism and Magnetic Materials}\
  }\textbf {\bibinfo {volume} {449}},\ \bibinfo {pages} {500} (\bibinfo {year}
  {2018})}\BibitemShut {NoStop}%
\end{thebibliography}%

\end{document}